\documentclass{pasj00}
\color{black}

\begin{document}
\SetRunningHead{Y. Ishisaki et al.}
{Suzaku XRT/XIS Simulator \& ARF Generator} 
\Received{2006/09/06}%{yyyy/mm/dd}
\Accepted{2006/09/29}%{yyyy/mm/dd}

\title{
Monte-Carlo Simulator and Ancillary Response Generator
of Suzaku XRT/XIS System for Spatially Extended Source Analysis
%Suzaku XRT + XIS Simulator and Ancillary Response Generator
%for Spatially Extended Source Analysis
}

\author{%
Yoshitaka \textsc{Ishisaki},\altaffilmark{1}
Yoshitomo \textsc{Maeda},\altaffilmark{2}
Ryuichi \textsc{Fujimoto},\altaffilmark{2}
Masanobu \textsc{Ozaki},\altaffilmark{2}\\
Ken \textsc{Ebisawa},\altaffilmark{2}
Tadayuki \textsc{Takahashi},\altaffilmark{2}
Yoshihiro \textsc{Ueda},\altaffilmark{3}
Yasushi \textsc{Ogasaka},\altaffilmark{4}\\
Andrew \textsc{Ptak},\altaffilmark{5}
Koji \textsc{Mukai},\altaffilmark{6}
Kenji \textsc{Hamaguchi},\altaffilmark{6}
Masaharu \textsc{Hirayama},\altaffilmark{6}
Taro \textsc{Kotani},\altaffilmark{7}\\
Hidetoshi \textsc{Kubo},\altaffilmark{8}
Ryo \textsc{Shibata},\altaffilmark{4}
Masatoshi \textsc{Ebara},\altaffilmark{2}
Akihiro \textsc{Furuzawa},\altaffilmark{4}
Ryo \textsc{Iizuka},\altaffilmark{9}
Hirohiko \textsc{Inoue},\altaffilmark{2}\\
Hideyuki \textsc{Mori},\altaffilmark{8}
Shunsaku \textsc{Okada},\altaffilmark{2}
Yushi \textsc{Yokoyama},\altaffilmark{2}
Hironori \textsc{Matsumoto},\altaffilmark{8}
Hiroshi \textsc{Nakajima},\altaffilmark{8}\\
Hiroya \textsc{Yamaguchi},\altaffilmark{8}
Naohisa \textsc{Anabuki},\altaffilmark{10}
Noriaki \textsc{Tawa},\altaffilmark{10}
Masaaki \textsc{Nagai},\altaffilmark{10}
Satoru \textsc{Katsuda},\altaffilmark{10}\\
Kiyoshi \textsc{Hayashida},\altaffilmark{10}
Aya \textsc{Bamba},\altaffilmark{11}
Eric D.~\textsc{Miller},\altaffilmark{12}
Kousuke \textsc{Sato},\altaffilmark{1}
Noriko Y.~\textsc{Yamasaki}\,\altaffilmark{2}
}
\altaffiltext{1}{
Department of Physics, Tokyo Metropolitan University, %\\
1-1 Minami-Osawa, Hachioji, Tokyo 192-0397}
\email{ishisaki@phys.metro-u.ac.jp}
\altaffiltext{2}{
Institute of Space and Astronautical Science (ISAS), JAXA, %\\
3-1-1 Yoshinodai, Sagamihara, Kanagawa 229-8510}
\altaffiltext{3}{
Department of Astronomy, Kyoto University, Sakyo-ku, Kyoto 606-8502}
\altaffiltext{4}{
Department of Particle and Astrophysics, Nagoya University, %\\
Furo-cho, Chikusa-ku, Nagoya 464-8602}
\altaffiltext{5}{
Department of Physics and Astronomy, Johns Hopkins University,\\
3400 North Charles Street, Baltimore, MD 21218-2686, USA}
\altaffiltext{6}{
NASA/Goddard Space Flight Center, Greenbelt, MD 20771, USA}
\altaffiltext{7}{
Department of Physics, Tokyo Institute of Technology,
2-12-1 O-okayama, Meguro-ku, Tokyo 152-8551}
\altaffiltext{8}{
Department of Physics, Kyoto University, Sakyo-ku, Kyoto 606-8502}
\altaffiltext{9}{
Nishi-Harima Astronomical Observatory, Sayo-cho, Hyogo 679-5313}
\altaffiltext{10}{
Department of Earth and Space Science, Osaka University, %\\
1-1 Machikane-yama, Toyonaka, Osaka 560-0043}
\altaffiltext{11}{
The Institute of Physical and Chemical Research (RIKEN), %\\
2-1 Hirosawa, Wako, Saitama 351-0198}
\altaffiltext{12}{
Kavli Institute for Astrophysics and Space Research, %\\
Massachusetts Institute of Technology, Cambridge, MA 02139, USA}
%\altaffiltext{6}{
%Shonan Institute of Technology, 1-1-25, Tsujidonishikaigan,
%Fujisawa-shi, Kanagawa}

\maketitle

\begin{abstract}

We have developed a framework for the Monte-Carlo simulation of the X-Ray
Telescopes (XRT) and the X-ray Imaging Spectrometers (XIS) onboard
{\sl Suzaku}, mainly for the scientific analysis of spatially 
and spectroscopically complex celestial sources.
A photon-by-photon instrumental simulator is built
on the ANL platform, which has been successfully used
in {\sl ASCA} data analysis.
The simulator has a modular structure,
in which the XRT simulation is based on a ray-tracing library,
while the XIS simulation utilizes a spectral
``Redistribution Matrix File'' (RMF), generated separately by other tools.
Instrumental characteristics and calibration results,
e.g., XRT geometry, reflectivity, mutual alignments,
thermal shield transmission, build-up of the contamination on the XIS
optical blocking filters (OBF),
are incorporated as completely as possible.
Most of this information is available in the form of the
FITS (Flexible Image Transport System) files in the standard
calibration database (CALDB). This simulator can also be utilized to
generate an ``Ancillary Response File'' (ARF),  which describes the
XRT response and the amount of OBF contamination.  The ARF is dependent
on the spatial distribution of the celestial target and the photon
accumulation region on the detector, as well as observing conditions
such as the observation date and satellite attitude.
We describe principles of the simulator and the ARF generator,
and demonstrate their performance in comparison with in-flight data.
\end{abstract}

\section{Introduction}

A Monte-Carlo simulator is useful in characterizing a detector,
and can relatively easily take into account many of the parameters
which affect observations. Since the ultimate goal of X-ray
data analysis is to estimate the true time, energy and position of the
incoming X-ray photons, it is quite important
to predict precisely how the photons interact with
the telescope and detector.
A good simulator is therefore strongly required not only
for instrumental calibration and proposal planning, but also for
scientific analysis.  
{\sl Chandra} and {\sl XMM-Newton} also have good simulators,
named MARX\,\footnote{http://space.mit.edu/ASC/MARX/}
and SciSim\,\footnote{http://xmm.vilspa.esa.es/scisim/},
respectively.
%It is also desirable for the simulator to be configurable
%and reusable in parts, so that instrumental team,
%software developers and end users can share software parts
%for commonly requested tasks with the latest calibration results.

The X-ray observatory {\sl Suzaku} (formerly known as {\sl Astro-E2}\/)
is the fifth Japanese X-ray astronomy satellite \citep{Mitsuda2006}.
It has been developed under a Japan--US international collaboration,
and was launched on 2005 July 10.
Five X-Ray telescopes are present, sensitive to soft X-rays below
$\sim 10$~keV (XRTs; \cite{Serlemitsos2006}). At the foci of four of
the XRTs (XRT-I) are charge-coupled devices (CCD), known as the X-ray Imaging
Spectrometers (XIS; \cite{Koyama2006});
one (XRT-S) is combined with an X-ray calorimater known as the X-Ray
Spectrometer (XRS; \cite{Kelley2006}; XRS quit operation $\sim 1$ month after the launch).

The {\sl Suzaku} XRT is characterized by large collective
area and relatively short focal lengths,
compared with those of {\sl Chandra} and {\sl XMM-Newton}.  
In combination with these features, the low-earth orbit of {\sl Suzaku},
where the particle background is low and stable,
makes the non-X-ray background (NXB) of {\sl Suzaku}
much lower compared with {\sl Chandra} and {\sl XMM-Newton}.
% if normalized for the same sky area.
In addition, the XIS achieves good
spectral resolution, especially at the low energy range
below $\sim 1$~keV with the backside-illuminated (BI) CCD for XIS1.
The front-illuminated (FI) CCDs for XIS0, XIS2, and XIS3
exhibit about half of the NXB rate than XIS1
(and less at energies $\gtrsim 8$~keV), so they are complementary. 
Therefore, {\sl Suzaku} has a unique advantage for spectroscopic
observations of spatially extended sources \citep{Mitsuda2006}.

To achieve large collective area within the tight weight budget (1706~kg),
the {\sl Suzaku} XRT adopts the conical approximation of Wolter type I optics
with 175 layers of the thin-foil-nested reflectors per quadrant
\citep{Serlemitsos2006}. In return for the high throughput, it provides
a moderate imaging capability of 2$'$ half power diameter with a complex
point spread function (PSF), as well as the energy-dependent vignetting effects
common to X-ray telescopes.  In addition, there exists
spatially-dependent contamination on the optical blocking filters (OBF) of
the XIS \citep{Koyama2006}. These XRT and XIS characteristics often
make extended source analysis complicated,
so it is crucial to prepare a tool in order to precisely
evaluate the effect of complex telescope and detector responses.

%A Monte-Carlo simulator is a powerful tool to characterize the response,
%and provides a straightforward way to take it into account in the
%analysis, once the parameters are determined. Therefore, it is very
%useful not just for the instrumental calibration and the proposal
%planning, but also for the scientific analysis.
We developed a Monte-Carlo simulator of the {\sl Suzaku} XRT/XIS system,
which is incorporated into two practical tools,
the XIS simulator ``{\sl xissim}'' and the ``Ancillary Response File'' (ARF)
generator ``{\sl xissimarfgen}''.
The simulator is constructed on the ``ANL'' platform (\S\,\ref{subsec:anl}),
which is used for almost all of the processing and analysis software of
{\sl Suzaku}.
%The ANL platform has a modular structure, such that 
%it is configurable and reusable in components.  Consequently,
%the instrument calibration teams, software developers,
%and end-users can share the same software components for different purposes.

While these tasks provide vast flexibility to the {\sl Suzaku} XIS users,
it is rather difficult to utilize them efficiently and appropriately.
For example, there are more than 90 parameters for both
{\sl xissim} and {\sl xissimarfgen}.
There are several issues and limitations that one should be aware of
in running these tasks.
This paper is aimed to clarify these things by explaining
principles of the software and by demonstrating performance
with practical examples.
%This paper is aimed to guide the users of {\sl
%xissim} and {\sl xissimarfgen}, as well as to introduce 
%useful functions and/or restrictions of the programs.
%We have intended to be clear and comprehensive as much as we can,
%so that one can find his/her solution in trouble by him/herself,
%although it might be a little bit verbose.
We have also tried to separate the `calibration issues',
which can be changed (usually improved) by calibration updates,
from those originated in the design of the software itself.
The quality of the calibration is out of scope for this paper,
although some aspects are discussed briefly in \S\,\ref{sec:demonstration}.

This paper is organized as follows.
In \S\,\ref{sec:suzaku-software}, we briefly show the strategy of the
{\sl Suzaku} software development,
focusing on the ANL platform and simulators.
In \S\,\ref{sec:xissim} and \S\,\ref{sec:xissimarfgen},
we describe principles of {\sl xissim} and {\sl xissimarfgen}, respectively.
In \S\,\ref{sec:discussion},
several notes on these tasks are described.
In \S\,\ref{sec:demonstration},
we demonstrate these tasks with three distinct examples, the
Crab Nebula, the North Ecliptic Pole (NEP) field, and Abell~1060.
Finally, a summary is given in \S\,\ref{sec:summary}.
We also added three appendices which describe
the coordinates definition, structures, parameters, and
the output file formats, in detail.

\section{Software Development for Suzaku}\label{sec:suzaku-software}

\subsection{The ANL Platform}\label{subsec:anl}

When {\sl ASTRO-E} software development started in 1995,
the goal was a common software framework/platform which is used by
realtime quick-look, data processing, and scientific analysis, both
during pre-launch phase and after the launch.  To that end, it was
necessary to provide a common programming environment where instrument
team members can easily develop, maintain and updates softwares
that they need.  This framework/platform also must allow end-users
to share these softwares.
The framework must be easy to learn for instrument team members,
who, spending most of the time in calibrating the instruments,
do not necessarily have extensive programming experience.
Also, from the end users' point of view, it is desirable
that those software tools developed based on this framework are
maximally flexible and have an {\sc Ftools}-like simple interface
which is familiar to most X-ray astronomers.

A software platform called ``ASCA\_ANL'',
which had been developed for the ASCA satellite \citep{Tanaka1994},
fulfills these  requirements.
The ASCA\_ANL platform mandates modular design of the analysis
software to be built upon it, and makes the software products easily
configurable and reusable in components, so that software developers
and end-users can share the same components for different purposes.
This feature not only reduces code duplication, 
but also helps to quickly mature and refine the software.  
% It is easy for the instrument teams to include newly discovered,
% unanticipated issues into the
% existing simulators, while configurable structure gives
% flexibility for end users.  

Indeed, ASCA\_ANL fostered many practical tools including the
instrument simulator SimASCA and the response generator SimARF\@.
The advantages of the ASCA\_ANL platform are demonstrated
by original scientific research which would have been difficult
without SimASCA and SimARF;
e.g., spatial-spectral analysis of clusters of galaxies
\citep{Ikebe1996,Honda1996},
systematic analysis of large volumes of X-ray surveys and the cosmic X-ray
background \citep{Ueda1998,Ueda1999,Kushino2002}.
The SimASCA and SimARF were very helpful to realize
a specialized analysis method  in the analysis of spatially extended
sources (which is not  supported by standard analysis software), 
and to accurately compute complicated instrument responses.

On the other hand, however,
ASCA\_ANL and other relevant software were based on the
functions and libraries used for realtime quick-look software
which had been developed by the instrument teams,
independently from the official {\sl ASCA} analysis software ({\sc Ftools}).
This resulted in two independent streams to calculate
basic physical values from the raw data, such as
the pulse height corrected for the detector gain changes
(known as ``pulse invariant'', or PI,
corresponding to the detected photon energy),  
and the sky and detector coordinates of events, which
caused confusion in the scientific analysis of the {\sl ASCA} data.

Based on the {\sl ASCA} experience,
we adopted the ``ANL'',  i.e.\ a generalized version of the ASCA\_ANL,
as the software development platform for {\sl ASTRO-E} and {\sl Suzaku}. 
A brief history and concept of the ANL are described in \citet{Ozaki2006}.
At the same time, we developed a mechanism to convert ANL software
directly to {\sc Ftools}, to ensure that the ANL tools used
for calibration by instrument teams are equivalent to {\sc Ftools}
used for pipeline processing and scientific data analysis.
Common FITS-read and -write ANL modules and functions were also developed
to handle photon event files and the calibration files in FITS
format.\footnote{ 
The calibration FITS files are released to
the public from the NASA/GSFC guest observer facility,
as part of the official calibration database
(CALDB; \cite{George1991}).}
Now, almost all of the {\sc Ftools} for {\sl Suzaku}
including {\sl xissim} and {\sl xissimarfgen}, 
released from the Guest Observer Facility at NASA/GSFC,
are developed in the ANL framework.

\subsection{History of Suzaku Simulators}

The development of the {\sl Suzaku} simulator had started
before the failure launch of the {\sl ASTRO-E} on 10 Feb 2000,
especially for the bilinear $16\times 2$ pixel XRS detector
\citep{Kelley1999}. The detector size was comparable
with the angular resolution of the XRT \citep{Kunieda2001,Shibata2001}, 
and so the XRS PSF was undersampled.
The energy resolution of the XRS was also very dependent
on the count rate of each calorimeter pixel.
Therefore, the XRT/XRS system simulator, {\sl xrssim},
was required to estimate the flux coming to each XRS pixel,
which was critical for proposal planning.
This was also the case for the Suzaku XRS\@.
%, although the Suzaku XRS also has been lost.
The {\sl xissim} task subsequently developed by replacing the XRS
component of the simulator with the XIS component.

The XRT ray-tracing part of the simulator has been significantly
updated from the {\sl ASCA} era.
The code had been rewritten, by R.~L.~Fink (NASA/GSFC),
from Fortran into C++,
and the structure had been re-designed to utilize
the mirror geometry and reflectivity files
as separate calibration FITS files.
This ray-tracing code is now supplied as the ``{\sl xrrt}\/'' library
by the XRT team. It has been utilized for the performance improvement
of the XRT \citep{Misaki2005} and the design of the pre-collimator to
suppress the stray-light \citep{Mori2005}.

At present, {\sl xissim} and {\sl xissimarfgen} are publicly released
in the {\sl Suzaku} {\sc Ftools} and all the source code is available
including the ANL itself and the {\sl xrrt}\/ library.
The latest version of the {\sl xissim} package is 2006-08-26,
which will be included in the next official release of
the {\sl Suzaku} {\sc Ftools} for version 2.0 processing of
{\sl Suzaku} archival data scheduled in late 2006.
All the calibration information currently available is taken into account
via the CALDB calibration database.
The {\sl mkphlist}, {\sl xissim}, {\sl xissimarfgen}, and
{\sl xiscontamicalc} tasks described in this paper
are based on this version of the {\sl xissim} package.
The latest information on the {\sl xissim} package is available at
http://www-x.phys.metro-u.ac.jp/\~\/ishisaki/xissim/.

We also note that there is another Monte Carlo simulator
for {\sl Suzaku}, based on the Geant4 toolkit \citep{Geant4_2003}
with ANL++ \citep{Ozaki2006}.
This can simulate interactions of
cosmic rays (both X/$\gamma$-rays and particles)
with the satellite materials,
such as satellite structures, shielding around detectors,
and the detectors themselves.
The main purpose of this simulator is to study
response of the Hard X-ray Detector (HXD; \cite{Takahashi2006,Kokubun2006}),
and the NXB models for both HXD and XIS\@.
See \citet{Terada2005} and \citet{Ozaki2006} for details.

%We thus started developing the Suzaku XRT + XIS simulator with the ANL
%platform, generalized ASCA\_ANL and now the backbone of the Suzaku
%public software FTOOLS, distributed as a part of HEAsoft,
%http://heasarc.gsfc.nasa.gov/lheasoft/. The simulator is called
%``xissim'' while the ancillary response generator is ``xissimarfgen''.
%Both softwares are assisted with a software ``mkphlist'' to support the
%analysis of extended cerestical objects. The softwares are released as a
%part of the Suzaku ftools. The reference files were distributed as the
%Suzaku CALDB. 

\section{Simulator: xissim}\label{sec:xissim}

The {\sl xissim} task simulates 
the interaction of the incident X-ray photons with
the XRT/XIS system, using the XRT ray-tracing library and
a spectral ``Redistribution Matrix File''
(RMF; see \S\,\ref{subsec:principle-and-limitations}) for
the XIS, and generates a simulated event file.
The format of the generated event file
is a stripped-down version of that created by the pipeline processing
of a real observation,
so that users can analyze the simulated data in the same manner as the
real data. 
To perform the simulation, users need to take three steps.
First, the spatial distribution and the energy spectrum of
the celestial source to be simulated must be specified.
Second, a list of incident photons from the source
needs to be prepared as FITS file(s).
An auxiliary tool, {\sl mkphlist}, may be used for this purpose.
Third, the photon FITS file is passed to {\sl xissim},
which then performs a photon-by-photon simulation,
and creates a file of events detected by the XIS\@.
In the following subsections, we describe how {\sl xissim}
performs the simulation.

\subsection{Photon Generation}

An auxiliary task {\sl mkphlist} generates a list of faked photons of an
X-ray source from the model spectral energy and spatial distribution,
photon flux (in photons~cm$^{-2}$~s$^{-1}$) in an arbitrary energy band,
and the geometrical area of the XRT (in cm$^2$), provided by users.
A model spectral distribution file
(which is specified by the {\sf qdp\_spec\_file} parameter)
must be in units of photon flux (photons~cm$^{-2}$~s$^{-1}$~keV$^{-1}$)
that can be easily produced with standard software packages such as
{\sc Xspec} \citep{Arnaud1996}.
%A photon file for an extended source can be generated
%by supplying an image of the source.
{\sl mkphlist} requires celestial coordinates of the point source
or a surface brightness map (FITS image) on the sky
for the spatial distribution of the source.
Either the number of photons or exposure time is needed
to determine how many photons are to be generated.
Users can also specify equal or random interval steps
for the photon arrival time.
The structure of {\sl mkphlist} is explained
in Appendix~\ref{subsec:mkphlist-structure},
and a list of major parameters and the format of the photon file
are summarized in table~\ref{tab:mkphlist}~and~\ref{tab:photon-file},
respectively.
Note that, by preparing an appropriate photon file,
users can in principle simulate any source with any energy spectrum
and/or any spatial distribution.

\subsection{Photon-by-Photon Simulation}\label{sec:photon-by-photon}

\begin{figure}[tbg]
\centerline{
\FigureFile(0.48\textwidth,0.48\textwidth){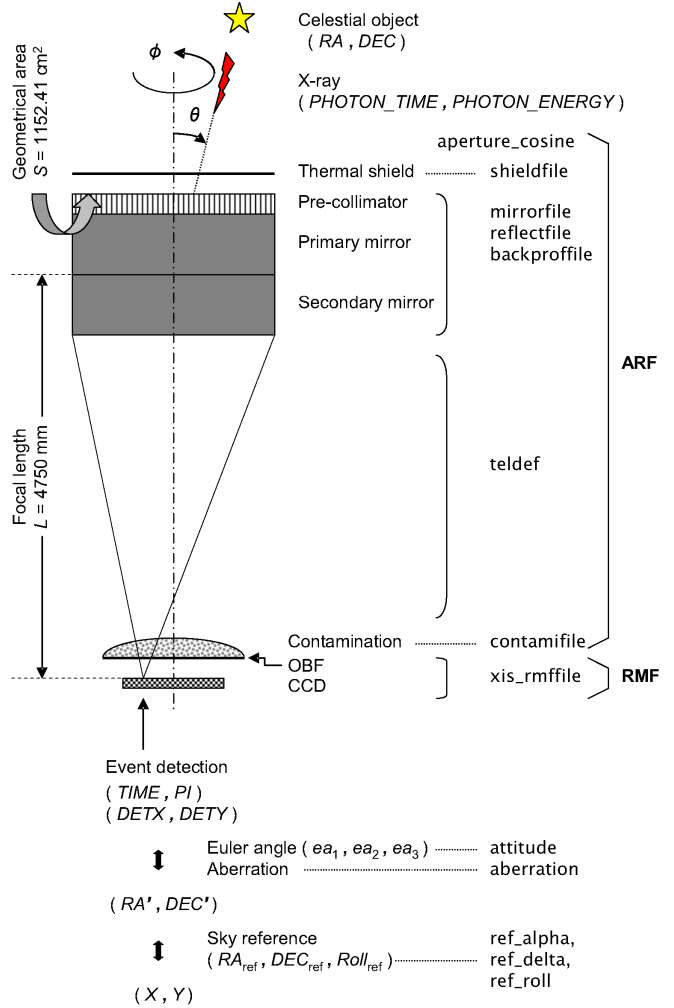}
}
\caption{
Schematic structure of the simulation.
}\label{fig:simulator-structure}
\end{figure}

By taking into account the XRT and XIS response,
{\sl xissim} performs photon-by-photon simulation
for given input photon file(s).
It has the capability to read up to eight photon files simultaneously.
Figure~\ref{fig:simulator-structure} shows the schematic structure
of the simulation implemented in {\sl xissim}.
Since understanding the coordinate systems is essential,
we include the definitions in Appendix~\ref{app:coordinates}.

First of all, the {\sc ra} and {\sc dec} values in the photon file
need to be converted to ($\theta$, $\phi$),
i.e.\ offset angle ($'$) and azimuth angle ($^\circ$),
with respect to the XRT optical axis.
This requires the satellite Euler angles
(${\it ea}_1$, ${\it ea}_2$, ${\it ea}_3$) ($^\circ$),
the observation date for an aberration correction
(or parallax correction, see Appendix~\ref{app:coordinates}),
and the alignment parameters in the telescope definition ({\it teldef}\/) file.
Users can supply an attitude file (set of Euler angles
as a function of time) and a good time interval (GTI) file
to take into account the wobbling of the spacecraft
(See also \S\,\ref{subsec:notes-accum-region} for the attitude wobbling).
The {\sc photon\_time} column in the photon file usually starts
from 0.0~s unless otherwise specified,
and it is treated as the time offset relative to the GTI\@.
Alternatively one may specify a fixed set of Euler angles and/or a fixed date.
The aberration correction can be disabled by setting
the hidden parameter {\sf aberration}=no (hidden parameters are not
required when invoking an {\sc Ftools} task).

In the second stage, the geometrical area for a given
photon is reduced by a factor of $\cos\theta$ due to
the slanted incidence to the XRT\@.
This factor is usually very close to unity,
and had been neglected in the older version of {\sl xissim}.
This behavior can be controlled with the parameter
{\sf aperture\_cosine}, and is set to `yes' by default in the present version.
The photon flux is further reduced due to transmission through
the thermal shield on the top of the XRT\@.
{\sl Xissim} then assigns a random location for each photon at
the top surface of the XRT, where the pre-collimator is placed.
The task traces the path of each photon inside the XRT
(pre-collimator, primary and secondary mirrors), using
the XRT ray-tracing library, {\sl xrrt} \citep{Misaki2005,Mori2005},
using the XRT geometry and reflectivity as described in the
ray-tracing code and the calibration files. 
After the ray-tracing, some photons may be absorbed and disappear,
while others reach the focal plane.

A fraction of the photons that have reached the focal plane are absorbed by the
contamination on the OBF. 
The thickness of the contamination is
time- and detector-position-dependent \citep{Koyama2006},
and their dependence is given by a calibration file
supplied by the XIS team. {\sl Xissim} computes the transmissivity
at a given time and position using this calibration file. 
The position of the photon on the detector is again calculated by
the alignment parameters in the {\it teldef}\/ file.

Finally, the simulated photons reach the detector
(including both OBF and CCD), where the detection
probability is determined using the RMF of the XIS\@.
The XIS RMF contains the transmission of the OBF
and the quantum efficiency of the CCD,
as well as the spectral redistribution matrix from energy to PI\@.
The line response function of the XIS CCDs is primarily a Gaussian
distribution but it also includes other features such as escape ratios
and tails that deviate from a Gaussian.
Photons that have passed the test for detection
are recorded as X-ray events,\footnote{
We shall call `photon' during the simulation,
which becomes `event' after the detection.}
and their PI values are determined
from the incident photon energy by random choices
according to the energy redistribution probability in the RMF\@.
The Suzaku XIS detectors do not exhibit significant positional
dependence in the energy resolution after the CTI
(charge transfer inefficiency) correction while the energy resolution
is known to degrade with time.
Users should supply an appropriate RMF corresponding to
the observation date, which can be generated by a separate task,
{\sl xisrmfgen}.

Note that the current version of {\sl xissim}
does not consider the NXB, bad CCD columns, event pile-up,
event grade, nor CCD exposure frames. Although the output event files contains
the same major columns as the event files of the real data, the {\sc status}
and {\sc grade} columns are filled with 0,
while the {\sc pha} column has the same value as the {\sc pi} column.

The ANL module structure and parameters of {\sl xissim} is
explained in Appendix~\ref{subsec:xissim-structure} in detail.
If one has the ANL programing environment available,
he/she may add his/her own modules to the simulator. 
It is also easy to replace a module,
e.g.\ if a module is available that more precisely simulates the CCD
detection process, then this module can be substituted for the
SimASTE\_XISRMFsim module which utilizes a ready-made RMF\@.
This is one of the great benefits of the ANL\@.  

\subsection{Calibration Files}

Table~\ref{tab:caldb} summarizes the list of
calibration files used by {\sl xissim}.
The file specified by the {\sf leapfile} parameter is
the leap second file, and is required to compute the mission time
(or {\sl Suzaku} time), defined as accumulative seconds
since 2000 January 1, 00:00:00 (UTC)\@.
In fact, the default value of the {\sf leapfile} parameter is
set to a special keyword of ``{\sc caldb}'' and is a hidden parameter.
By installing CALDB
and properly setting the environmental variables,
the {\sl xissim} task automatically searches the most recent
calibration file for this category, i.e.\ `Content Name' = {\sc leapsecs}.
This is also applicable to other parameters in table~\ref{tab:caldb}.

The {\sf shieldfile}, {\sf mirrorfile}, {\sf reflectfile},
and {\sf backproffile} parameters are used for the XRT
simulation. There are four FITS extensions in the {\sf mirrorfile} 
to describe the geometry of each XRT, and three extensions are present
in the {\sf reflectfile} corresponding to materials of the reflection surface.
As described in Appendix~\ref{app:coordinates},
{\sf teldef} is used to describe the mutual alignments
between XRT and XIS, as well as among the XIS sensors and the spacecraft.
The {\sf contamifile} describes the energy, time, and position
dependence of the contamination on the XIS OBF\@.

\begin{table*}
\caption{
List of calibration files used by {\sl xissim}
}\label{tab:caldb}
\centerline{\small
\begin{tabular}{llll}
\hline\hline
Parameter	& File Name$\;^*$ & Content Name$\;^\dagger$ & Description \\
\hline
{\sf leapfile}	& {\tt leapsec\_010905.fits}
& {\sc leapsecs}
&	Table of times at which leap seconds occurred \\
{\sf shieldfile} & {\tt ae\_xrta\_shield\_20061129.fits}
& {\sc ftrans}		&	XRT thermal shield transmission \\
{\sf mirrorfile} & {\tt ae\_xrt{\it N}\_mirror\_20060710.fits}
& {\sc geometry}	&	XRT mirror geometry \\
&
& {\sc geometry}	&	XRT obstruction geometry \\
&
& {\sc geometry}	&	XRT quadrant geometry \\
&
& {\sc geometry}	&	XRT pre-collimator geometry \\
{\sf reflectfile} & {\tt ae\_xrta\_reflect\_20060710.fits}
& {\sc reflectivity}	&	XRT mirror foil front surface reflectivity \\
&
& {\sc reflectivity}	&	XRT mirror foil back surface reflectivity \\
&
& {\sc reflectivity}	&	XRT pre-collimator surface reflectivity \\
{\sf backproffile} & {\tt ae\_xrta\_backprof\_20060719.fits}\hspace*{-0.5em}
& {\sc backprof}	&	XRT foil backside scattering profile \\
{\sf teldef} & {\tt ae\_xi{\it N}\_teldef\_20060125.fits}
& {\sc teldef}	&	Telescope definition file \\
{\sf contamifile} & {\tt ae\_xi{\it N}\_contami\_20060525.fits}
& {\sc contami\_growth}\hspace*{-0.5em}
&	XIS OBF contamination growth curve \\
&
& {\sc contami\_trans}
&	Template transmission vs energy for the contaminant \\
\hline\\[-1ex]
\multicolumn{4}{l}{\parbox{0.96\textwidth}{\footnotesize
\footnotemark[$*$] $N$ represents 0, 1, 2, or 3 respective to the XIS sensor.
}}\\
\multicolumn{4}{l}{\parbox{0.96\textwidth}{\footnotesize
\footnotemark[$\dagger$] The CCNM{\it nnnn} keyword in the FITS header,
and the CAL\_CNAM column in the CALDB index file.}}
\end{tabular}
}
\end{table*}

\section{Ancillary Response Generator: xissimarfgen}\label{sec:xissimarfgen}

The {\sl xissimarfgen} task generates a {\sl Suzaku} XIS ARF
based on user-defined conditions,
such as an arbitrary shape of the X-ray emitting region
and event extraction regions.
{\sl Xissimarfgen} does so by simulating photon detections at each
energy. It then calculates the detection efficiency 
in a user-defined event accumulation region.\footnote{
We shall use `accumulate' for the simulated events,
and `extract' for the real observed events.}
Since it utilizes a Monte-Carlo simulation,
users need to simulate enough photons to avoid
counting statistics errors. It can refer to
the attitude file to reflect the change of effective area
due to the attitude wobbling.
The final ARF is in the standard FITS format, so that users can use {\sc Xspec}
or other standard fitting packages for spectral analysis.

\subsection{Principle of ARF Calculation and Limitations}
\label{subsec:principle-and-limitations}

The ARF is utilized for spectral fitting combined with an RMF\@.
See \citet{George1992} for detailed format of these files.
The RMF is represented by an $(m\times n)$ matrix $R(E_i,{\it PI}_j)$,
where $E$ (keV) denotes the energy and {\it PI} (channel; hereafter
chan) denotes the 
pulse invariant, with $1\le i\le m$ and $1\le j\le n$.
Regarding the XIS, $m=7900$, $n=4096$,
$E_1 = 0.201$~keV, $E_m = 15.999$~keV,
${\it PI}_1 = 0$~chan, and ${\it PI}_n = 4095$~chan
for the nominal RMF\@.
The ARF is represented by an $m$-dimensional vector which we denote as
$S\;A(E_i)$ (cm$^2$),
where $S=1152.41$~cm$^2$ represents the geometrical area of the XRT\@.
The goal of the spectral fitting is to find a model spectrum,
$M(E_i)$ (photons~cm$^{-2}$~s$^{-1}$~keV$^{-1}$),
which fits the observed spectrum,
$\mathcal{D}({\it PI}_j)$ (count~cm$^{-2}$~s$^{-1}$).
%Because dimensions of $M(E_i)$ and $\mathcal{D}({\it PI}_j)$
%are different, 
The response and model spectrum are convolved, i.e., 
\begin{eqnarray}\label{eq:rmf-operation}
\mathcal{M}({\it PI}_j) =
S\; \sum_{i=1}^{m} \Delta E_i\; A(E_i)\; R(E_i,{\it PI}_j)\; M(E_i),
\end{eqnarray}
where $\Delta E$ (keV) is the energy bin width, and $\mathcal{M}({\it PI}_j)$ and $\mathcal{D}({\it PI}_j)$ are compared.
As one can see easily from this formula,
$\left[\; A(E_i)\; R(E_i,{\it PI}_j)\;\right]$ represents
an expected spectrum for the monochromatic X-ray of $E=E_i$~keV,
and $\displaystyle A(E_i)\; \sum_{j=1}^n R(E_i,{\it PI}_j)$
represents the detection efficiency at $E=E_i$~keV\@.

Thus, calculating the ARF is reduced to the computation of
the detection efficiency at each energy step, $E_i$, of the RMF\@, a
job well-suited for a Monte-Carlo simulation.
For a given input $N_{\rm in}$ counts of monochromatic X-ray photons at
$E=E_i$~keV, the simulator predicts $N_{\rm det}$ detected
events and then the detection efficiency is simply
$A(E_i) = N_{\rm det} / N_{\rm in}$.
However, one must be very careful because the detection efficiency,
namely $N_{\rm det}$, is influenced by many factors:
first of all, the accumulation region of the event on the detector,
and the spatial distribution of the celestial sources assumed on the sky.
It is also affected by the observational conditions, such as
the satellite Euler angles, the date of the observation due to the
thickness of the XIS contamination and the parallax correction, etc.
The quality of the calibration and/or the Poisson statistics
can also impact $N_{\rm det}$.
{\em It is therefore important that
one must reproduce the user-selection and the observational
conditions of the real data as much as possible in the simulation}.
One must also take care to perform a simulation such that the photon
statistics are sufficiently better than the statistics of the real observation.

In fact, the spatial distribution on the sky
is sometimes complex and/or extended on a scale larger than
the telescope FOV\@.  Thus the accuracy of the spatial model can become
a major cause of systematic error in the estimation of the
detection efficiency, which leads to uncertainty in the source flux.
For example, if one assumes a more core-concentrated image
than in reality, more photons will be simulated to arrive at the detector,
which will over-estimate the detection efficiency.
One can test the assumed spatial distribution on the sky
by comparing the real observation image and the simulated one.

There is also another limitation due to the spectral fitting
procedure itself. In the conventional spectral fitting package
(e.g., {\sc Xspec} v11 or before), one can choose only a single response
matrix (ARF + RMF) for an observed spectrum in the spectral fitting.\footnote{
This restriction no longer holds in the latest release of {\sc Xspec} v12,
which allows different model components to have their own response.}
For example, an observed spectrum may contain thermal emission
which obeys an oval surface brightness profile,
as well as the cosmic X-ray background (CXB) spectrum of a
$\Gamma\sim 1.4$ power-law which extends nearly uniformly on the sky.
The ARF response for the oval surface brightness
is different from that for the uniform-sky emission,
hence one cannot fit the observed spectrum with the thermal model
+ power-law model in a usual way.
Strictly speaking, the energy spectrum should be the same at every point
in the assumed spatial distribution on the sky in order to conduct
spectral fitting with a single ARF + RMF response.
%Such a case in the real world is a rare except for a point source
%or an almost uniform emission like the bright (sunlit) Earth emission.

\subsection{Implementation of ARF Calculation}
\label{subsec:xissimarfgen-implement}

As described in \S\,\ref{sec:photon-by-photon}
and shown in figure~\ref{fig:simulator-structure},
the XIS RMF takes care of the OBF transmission and
the quantum efficiency of the CCD, hence
the XIS ARF should consider other factors for the detection efficiency,
namely, the thermal shield transmission, XRT effective area,
transmission of the OBF contaminant, etc.
Detailed explanation of structure, parameters, and the output ARF format
are given in Appendix~\ref{app:structure}~and~\ref{app:file-formats}

It reads a number of parameters
which specify the simulation conditions
(table~\ref{tab:xissimarfgen}), and
(1) determines energy steps to calculate detection efficiency;
(2) generates monochromatic photons (or quasi-monochromatic
within the narrow energy range) 
until the user-specified condition on the photon statistics is
fulfilled at each energy step;
(3) conducts the ray-tracing simulation for each photon;
(4) counts up the number of detected events
at each energy; 
(5) records the detection efficiency at each RMF energy bin
to the output ARF(s) by interpolating the simulation result;
(6) continues to the next energy step and loops to step~(2).

Note that the energy step determined in step~(1) is usually not same
as the RMF energy bin, because the computation time would be very long
to conduct photon-by-photon simulations in standard XIS RMF 2~eV steps
up to 16~keV\@. 
Interpolation is therefore required in step~(5).
In addition, the XRT effective area usually changes only gradually
with energy except for several characteristic energies such as the Au-M, Au-L,
and Al-K edges,\footnote{The front surface of the XRT reflector is coated with
gold and its substrate is made of aluminum. The pre-collimator is
made of aluminum, too, so that the Al-K edge appears in
the large-offset-angle response of the XRT\@.}
so that we may often choose sparse energy steps.
This feature can save the computation time effectively.

In the calculation of the detection efficiency,
the three major factors of
(i) transmission of the XRT thermal shield,
(ii) effective area (cm$^2$) of the XRT, and
(iii) transmission of the XIS OBF contaminant
are treated separately.
They are also written in separate columns in the resultant ARF
as {\sc shield\_transmis}, {\sc xrt\_effarea}, and {\sc contami\_transmis}
(table~\ref{tab:arf-columns}).
The resultant detection efficiency times the geometrical area,
$S\;A(E_i)$ (cm$^2$), is written in the {\sc specresp} column,
i.e., {\sc specresp} = {\sc shield\_transmis}
$\times$ {\sc xrt\_effarea} $\times$ {\sc contami\_transmis}.
Note that (i) and (iii) are supplied in the calibration files
%specified by the {\sf shieldfile} and {\sf contamifile} parameters
(table~\ref{tab:caldb}) in fine energy steps of $\sim$ eV,
whereas (ii) is usually calculated in more sparse energy step.
By separating these factors, one can obtain a good quality ARF
even in a sparse energy step for the simulations,
and moreover, one may remove, scale, or multiply
the {\sc contami\_transmis} factor afterwards.
The {\sl xiscontamicalc} task is provided to do
this kind of the ARF manipulation.

Note that the thickness of the OBF contaminant is
positionally dependent. It is therefore required to know
the spatial distribution of photons\,\footnote{%
This distribution is approximated in {\sl xissimarfgen} by a DET
coordinate image binned prior to applying absorption due the XIS
contamination.} 
falling on the OBF
at each RMF energy bin in order to evaluate the {\sc contami\_transmis} factor.
This energy dependence of the photon distribution is also determined
by interpolation, which incurs additional calculation time when the simulation
energy step is much wider than the RMF energy bin.
It is not easy to estimate the true photon distribution from the real
observation data, because the observed image is affected by the XRT
vignetting and the OBF contaminant, both of which are energy dependent.
Vignetting is a more severe effect in the higher energy band,
and the OBF contamination is severe in the lower energy band.
In addition one must subtract background to utilize the observed image.
The combined energy and spatial dependence of the XIS contamination
is considered in the ARF generator rather than the RMF generator,
for this reason.

At each simulation energy, in fact, the $A(E)$ value
is calculated using the weighted sum of events,
$N_{\rm w}$, instead of $N_{\rm det}$, as,
\begin{eqnarray}\label{eq:detection-efficiency}
A(E) &=& N_{\rm w}(E) \;/\; N_{\rm in}(E)\; = \sum_{k=1}^{N_{\rm in}(E)} w_k(E) \;/\; N_{\rm in}(E),
\end{eqnarray}
in which $w_k(E)$ denotes the {\sc weight} value (see 
Appendix~\ref{subsec:xissim-structure}) of 
each simulated photon at the energy of $E$~keV\@.
As mentioned above, the resultant $A(E_i)$ values at the RMF energy bin
are calculated by interpolation, complicated somewhat by when
the transmission of the OBF contaminant is considered.
Here, we define $l\equiv \makebox{\sc index}_i$
(see Appendix~\ref{subsec:xissimarfgen-output} for $\makebox{\sc index}_i$).
There are $N_{\rm in}(E'_l)$ photons with {\sc weight}
without contamination represented by $w_k(E'_l)$ and energy
a little below $E_i$,
and $N_{\rm in}(E'_{l+1})$ photons with $w_{k'}(E'_{l+1})$,
energy a little above $E_i$,
i.e., $E'_l\le E_i\le E'_{l+1}$.
The transmission of the OBF contaminant is calculated for
each of the simulated photons, as $\tau_{\rm k}(E_i,\makebox{\sc
  photon\_time}_k, 
\makebox{\sc detx}_k,\makebox{\sc dety}_k)$.
Note that the energy of each simulated photon,
$E'_l = \makebox{\sc photon\_energy}_k$, has been
replaced by the energy of the RMF bin, $E_i$.
Thus {\sl xissimarfgen} computes the final detection efficiency, $A(E_i)$,
with contamination, as
\begin{eqnarray}
A(E_i) = 
\makebox{\sc s}_i \!\!
\sum_{k=1}^{N_{\rm in}(E'_l)}\!\!
\frac{\tau_k\; w_k}{N_{\rm in}(E'_l)} \;+\;
\makebox{\sc t}_i \!\!\!\!
\sum_{k'=1}^{N_{\rm in}(E'_{l+1})}\!\!
\frac{\tau_{k'}\; w_{k'}}{N_{\rm in}(E'_{l+1})}\;,
\end{eqnarray}
by an interpolation. The definitions of 
$\makebox{\sc s}_i$ and $\makebox{\sc t}_i$
are given in eqs.~(\ref{eq:s}) and (\ref{eq:t}).

It also calculates the relative error
of $A(E)$ at each simulation energy, and the interpolated values
are stored in the {\sc relerr} column of the output ARF
(table~\ref{tab:arf-columns})\@. This column is useful to judge
the photon statistics is sufficient for the ARF calculation.
The relative error is calculated as,
\begin{eqnarray}\label{eq:relerr}
\makebox{\sc relerr} = 
\sqrt{\frac{N_{\rm in} - N_{\rm det}}{N_{\rm in} N_{\rm det}}} =
\sqrt{\frac{1}{N_{\rm det}} - \frac{1}{N_{\rm in}}}\;,
\end{eqnarray}
if $N_{\rm in}\,N_{\rm det}\,(N_{\rm in} - N_{\rm det}) \neq 0$,
otherwise {\sc relerr} = 1.0.
The derivation of this formula is a little tricky, because we know the
detected count $N_{\rm det}$ and the undetected count
$N_{\rm in}-N_{\rm det}$ in the simulation,
and both are considered to follow the Poisson statistics.
Since $A(E)$ is expressed as
$A(E) = N_{\rm det} / N_{\rm in}
     = 1 - (N_{\rm in} - N_{\rm det})/N_{\rm in}$,
the error of $A(E)$ can be evaluated in two ways,
$\delta A_1 = \sqrt{N_{\rm det}} / N_{\rm in}$ or 
$\delta A_2 = \sqrt{N_{\rm in} - N_{\rm det}} / N_{\rm in}$.
We therefore defines the relative error as
$\delta A / A = 1 / \sqrt{ (\delta A_1)^{-2} + (\delta A_2)^{-2} } / A
= \sqrt{ (N_{\rm in} - N_{\rm det}) / N_{\rm in} / N_{\rm det} }$
= eq. (\ref{eq:relerr}).

\section{Notes}\label{sec:discussion}

In this section,
we describe several notes on {\sl xissim} and {\sl xissimarfgen}.
\S\,\ref{subsec:random} applies to both tasks,
and others apply mainly to {\sl xissimarfgen}.

\subsection{Notes on Random Numbers}\label{subsec:random}

The quality of the random number generator to be used can affect the
quality of the Monte-Carlo simulation results.
A good random number generator should include a
very long cycle, fast computation, and wide significant bits.
{\sl xissim} and {\sl xissimarfgen} use an internal random number generator
in the {\sl astetool}\/ library,\footnote{%
See also Appendix~\ref{app:coordinates} for the {\sl astetool}\/ library.}
utilized by all modules.
This generates double precision floating point values
in the range of $0\le r< 1.0$ based on the Tausworthe method
\citep{Tausworthe1965}.
The generated random number has 62 significant bits
($\simeq 4.6\times 10^{18}$)
and its cycle is estimated to be about $2^{250}\simeq 10^{75}$.
These parameters are significantly wider and longer than the usual
random number function, {\it int rand(void)}, implemented in the
standard C library. 

Its code is machine independent, and it reproduces exactly
the same series of random numbers as long as the {\sf rand\_seed}
and {\sf rand\_skip} parameters are the same.
It is recommended to set a prime number (except 2) to {\sf rand\_seed}
for good randomization. The default value of {\sf rand\_seed}
for the simulation tasks is 7. They record the number of
random numbers generated in the simulation to the output event file,
as the {\sc randngen} keyword in the FITS header.
One may re-continue the simulation with the same series of random numbers
by setting the {\sf rand\_skip} parameter to its value.
However this code is not multi-thread compliant,
which may need to be upgraded in the future for faster (i.e.,
distributed) simulations.

\subsection{Notes on Accumulation Region}\label{subsec:notes-accum-region}

There is a difference
between specifying the accumulation region in SKY coordinates
versus DET coordinates. This may be ignored only when
the attitude wobbling and the parallax correction are negligible.
The accumulation region is fixed on the CCD when it is specified
in DET coordinates.
On the other hand, it moves around the CCD when specified
in SKY coordinates, according to the attitude wobbling.
In both cases, the celestial target moves around the CCD and is affected
by the vignetting effect of the XRT, also due to the attitude wobbling.
{\sl Xissimarfgen} can treat both situations correctly,
as far as the supplied attitude file is reliable, 
so that one should select the {\sf region\_mode} parameter
to match the extraction method of the real observation spectrum.

It is known that there is an unexpected attitude wobbling of $\sim 0.5'$
due to thermal distortion of supporting structure \citep{Serlemitsos2006},
however this effect is not included in the present attitude file.
This situation will be improved in near future by a dedicated
{\sc Ftool}, the {\sl aeattcor} task.
Until then, it is recommended to avoid using
too small of an accumulation radius ($r\lesssim 3'$).
One may check this effect by changing the accumulation radius
and test whether the fit results are affected significantly.
Alternatively, one may track the position of the PSF core
on the CCD for bright point-like source targets.

It is also notable that the background files for the XIS currently released,
which are a collection of events when the XRT was pointed to
the night (non-sunlit) Earth, do not support event extraction in
SKY coordinates. This situation will be improved in the future.
One may extract the background from the outer ring of the target,
however, this region also contains the outskirt of the PSF of
the main target, CXB, and the instrumental background,
which have a small dependence on the detector position.
The former two effects can be evaluated by {\sl xissimarfgen},
and the latter can be tested with the released background file.

\subsection{Notes on Flux Normalization}

Because the detection efficiency defined in
eq.~(\ref{eq:detection-efficiency}) is considered for
all the input photons coming from everywhere
in the supplied source image,
the normalization of the flux in spectral fitting gives
the value integrated over the whole region of the source image.
Therefore, if one generates a uniform-sky ARF with
$\makebox{\sf source\_rmax}=20'$ to fit the CXB spectrum,
the fit gives the flux from the
$\pi\cdot \makebox{\sf source\_rmax}^2 = 1257$ arcmin$^2$ sky area,
then the user needs to divide the flux by this area
to convert it to a surface brightness.

Other cases can similarly be complex, e.g.,\ an analysis of
a cluster of galaxies. Extracting spectra from annular rings
centered on the cluster core is frequently performed in the cluster analysis.
Here, we assume that the cluster emission spectrum is identical
everywhere on the sky, and only the normalization of the flux
decreases as the distance from the cluster core increases.
We also assume that the spatial distribution of the cluster on the sky
can be perfectly predicted, which has been supplied to {\sl xissimarfgen}
as the source image. Then the fit results for each ring should give
the same flux, while the observed count per unit area
decreases as the ring radius increases,
since the flux for the whole cluster is calculated for each fit.

If one gets different fluxes for each fit, then
this is the 1st order approximation of the correction factor
to the assumed source image at each ring.
It is often desired to derive the flux only coming from each ring.
To help with this kind of task, there is a keyword,
{\sc source\_ratio\_reg}, written in the output ARF
(table~\ref{tab:arf-header-keys}). This keyword holds the ratio
of the source image inside the specified accumulation region for the ARF,
which has been calculated during the simulation.
By multiplying this factor by the obtained flux,
the user can calculate the flux in that ring.

\subsection{Notes on Computation Time and Memory}

The code of {\sl xissimarfgen} is designed to conduct
the computation of an ARF as efficiently as possible
in both time and memory, although it still requires 
a significant amount of both.
The simulation code has been tuned for speed;
it reads all the required information including the attitude
into memory before the simulation.
Searches of tables such as reflectivity, transmission,
spatial and spectral distributions are
accelerated by adding an index.
Several functions cache previous values to skip redundant
calculations especially when the photon energy and/or time
is similar to the previous ones.
In addition, the binary distribution of the {\sl xissim}/{\sl xissimarfgen}
package is compiled with fast C compilers
using the highest optimization option.

The required memory is usually around 130~MB,
hence recent machines can easily run
{\sl xissimarfgen} task in memory.
The actual calculation time is very dependent on the simulation
energy step, the photon statistics,
as well as the computer platforms.
Table~\ref{tab:arf-calc-time} shows examples of computation time
on the AMD Athlon$^{\makebox{\tiny\sc TM}}$64\, 2.4~GHz CPU
with a 64-bit Linux OS\@.
The example (A) is the ARF shown in \S\,\ref{subsec:crab},
in which full observational features are taken into account,
and a {\sl Chandra} image ($1800\times 1800$ pixels) was supplied for
the {\sf source\_image} with {\sf source\_mode\sc=skyfits}.
Parameters of {\sf num\_photon}=100000 and
{\sf estepfile\sc=sparse} (55 energy steps) were chosen,
so that $5.5\times 10^6$ photons were simulated for the ARF calculation.

The time difference between (A) and (B) indicates that
significant fraction of time ($\sim 70$~s) was consumed
in the calculation of the XIS contamination.
However, this time is only proportional to $m\times N_{\rm det}$
and does not depend on the simulation energy step,
hence it should be acceptable.
The parallax (aberration) correction also
needs the non-negligible cost of $\sim 9$~s,
which is proportional to $N_{\rm in} = \makebox{\sf num\_photon}$.
Similar time is needed for the randomization in
the spatial distribution of the Crab nebula,
as seen in (C) $\rightarrow$ (D)\@.
The consumed time in the XRTsim module is also displayed by the ANL,
and it was 75.9~s for (D)\@.
This indicates that the ray-tracing code can perform the simulation
of a single X-ray photon in less than 15~$\mu$s on this machine.

\begin{table}
\caption{
Examples of computation time.
}\label{tab:arf-calc-time}
\centerline{\small
\begin{tabular}{lrrl}\hline\hline
\multicolumn{2}{r}{CPU Time $^*\!$} &
               $\Delta t$ $^\dagger$ & Simulation Condition $^\ddagger$\\
\hline
(A) & 179.5 s & ---\ \ \  & Full, {\sf source\_mode\sc=skyfits} \\
(B) & 109.9 s & $-69.6$ s & $\rightarrow$ {\sf contamifile\sc=none} \\
(C) & 101.0 s & $ -8.9$ s & $\rightarrow$ {\sf aberration}=no \\
(D) &  92.3 s & $ -8.7$ s & $\rightarrow$ {\sf source\_mode\sc=j2000} \\
\hline\\[-1ex]
\multicolumn{4}{l}{\parbox{0.47\textwidth}{\footnotesize
\footnotemark[$*$]
Total number of CPU-seconds that the process used directly in user mode,
measured by the Linux {\sl time} command.}} \\
\multicolumn{4}{l}{\parbox{0.47\textwidth}{\footnotesize
\footnotemark[$\dagger$]
Difference of time compared with the preceding line.}} \\
\multicolumn{4}{l}{\parbox{0.47\textwidth}{\footnotesize
\footnotemark[$\ddagger$]
Parameters of {\sf num\_photon}=100000 and
{\sf estepfile\sc=sparse} (55 energy steps) are common to all.}}
\end{tabular}
}
\end{table}

\section{Demonstration}\label{sec:demonstration}

In this section,
we demonstrate how {\sl xissim} and
{\sl xissimarfgen} work using three distinct examples:
the Crab nebula as a calibration source
and a quasi-point-like source in \S\,\ref{subsec:crab},
the NEP field as a ``blank sky'' in \S\,\ref{subsec:nep}, and
Abell~1060 as an spatially extended source in \S\,\ref{subsec:a1060}.

\subsection{Crab Nebula}\label{subsec:crab}

\begin{figure*}[p]
\centerline{
\FigureFile(\textwidth,\textwidth){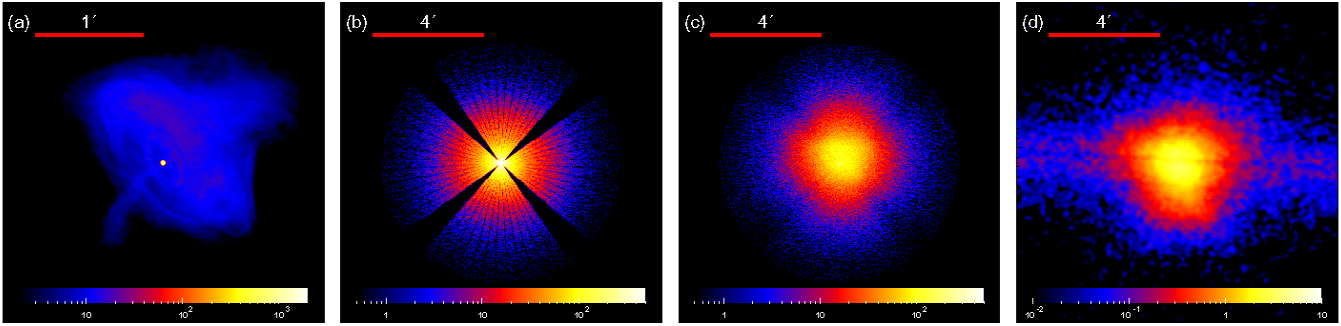}}
\caption{
Observed and simulated images of the Crab nebula. 
(a) {\sl Chandra} image \citep{Weisskopf2000}, corrected for pile-up.
(b) Simulated {\sl Suzaku} image for a point-like source.
(c) Simulated {\sl Suzaku} image using (a) as the spatial distribution.
(d) Observed {\sl Suzaku} XIS0 image smoothed with a Gaussian of
$\sigma=4$ pixel $\simeq 4''$.
The image width of {\sl Chandra} is 2.9$'$, while
those of {\sl Suzaku} are 11.6$'$. See text for details. 
}\label{fig:crab-image}
\end{figure*}

First, we present the case of simulating the Crab nebula
which is the main X-ray calibration source for effective area calibration
\citep{Toor1974,Seward1992,Kirsch2005}.
Small in angular scale, it has a complex spatial structure
as seen in figure~\ref{fig:crab-image}~(a) of
the {\sl Chandra} image \citep{Weisskopf2000}.
With respect to the surface brightness map,
we adopted this image, because {\sl Chandra}'s X-ray telescope,
HRMA, has much superior angular resolution of $\sim 0.5''$.
%almost two orders of magnitude better than that of {\sl Suzaku}.
We further compensated it manually for a point-like emission
from the neutron star (K.~Mori priv.\ comm.).

We made a photon list by supplying the image
to {\sl mkphlist}, and ran {\sl xissim} with it.
The simulated {\sl Suzaku} image of the Crab nebula is shown
in figure~\ref{fig:crab-image}~(c).
For comparison, we also present the simulated image for
a point-like source in figure~\ref{fig:crab-image}~(b).
The simulated Crab image appears as a smoothed PSF
with the extent of the complex surface brightness profile of the Crab nebula.
Figure~\ref{fig:crab-image}~(d) shows the real observation
image taken with the {\sl Suzaku} XIS0 detector.
The global extent of the Crab image is consistent with
that of the simulated image. The anisotropy in the azimuth
direction in the real image is due mainly to
the complex PSF shape of the actual XRT,
which will be more accurately reproduced by future improvement
of the mirror geometry file ({\sf mirrorfile}).
Once the calibration file is updated,
{\sl xissim} can reflect it automatically via the CALDB\@.
A narrow groove crossing the central area from east to west
is due to a bad CCD column.
The out-of-time events, which broadly spread on both the east and west
sides,
are also seen along the direction of the signal transfer
from imaging area to frame-store region.
These features are not implemented in the current version of {\sl xissim}.

\begin{figure*}[p]
\centerline{
\FigureFile(0.45\textwidth,0.48\textwidth){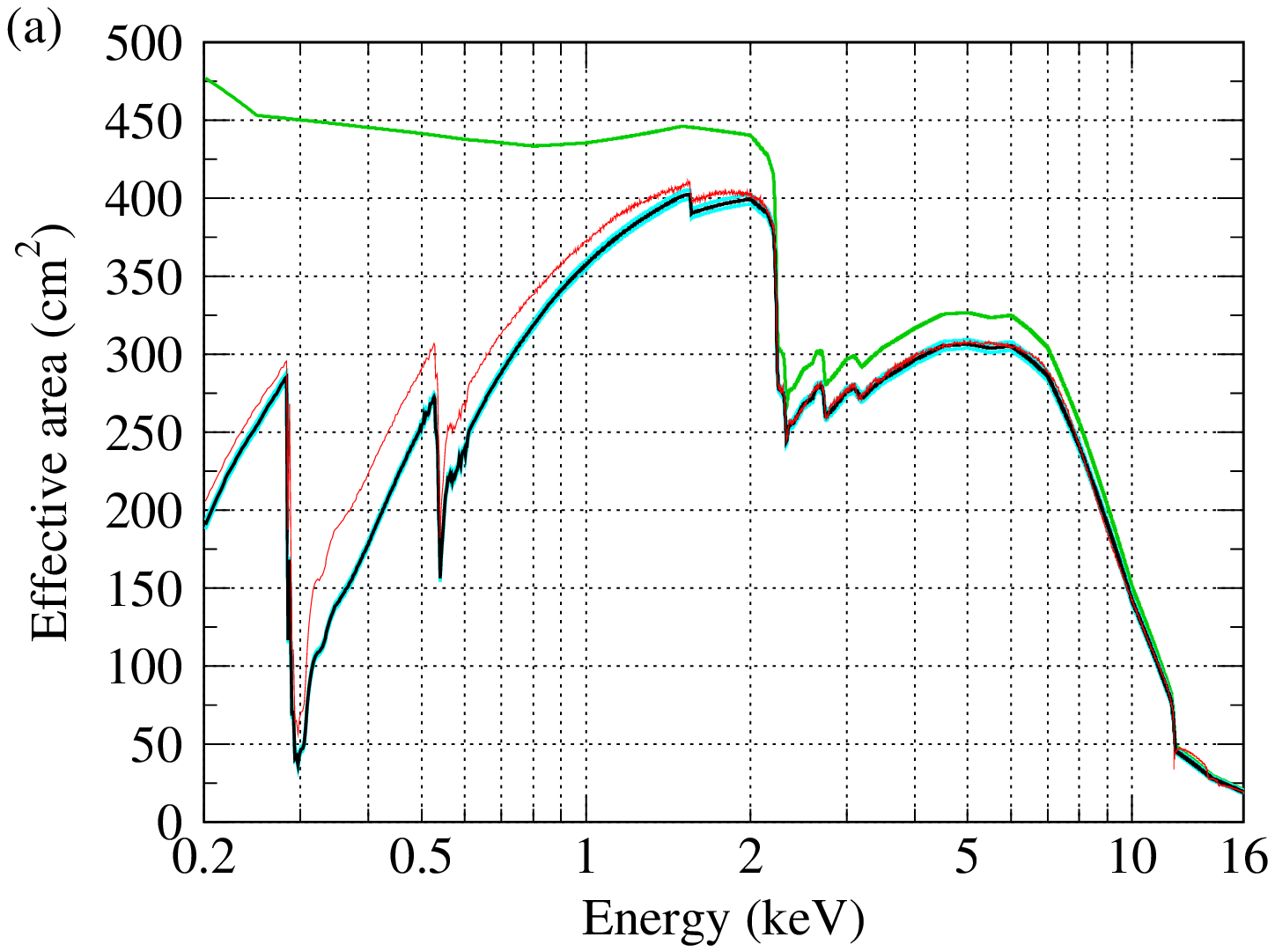}%
\FigureFile(0.45\textwidth,0.48\textwidth){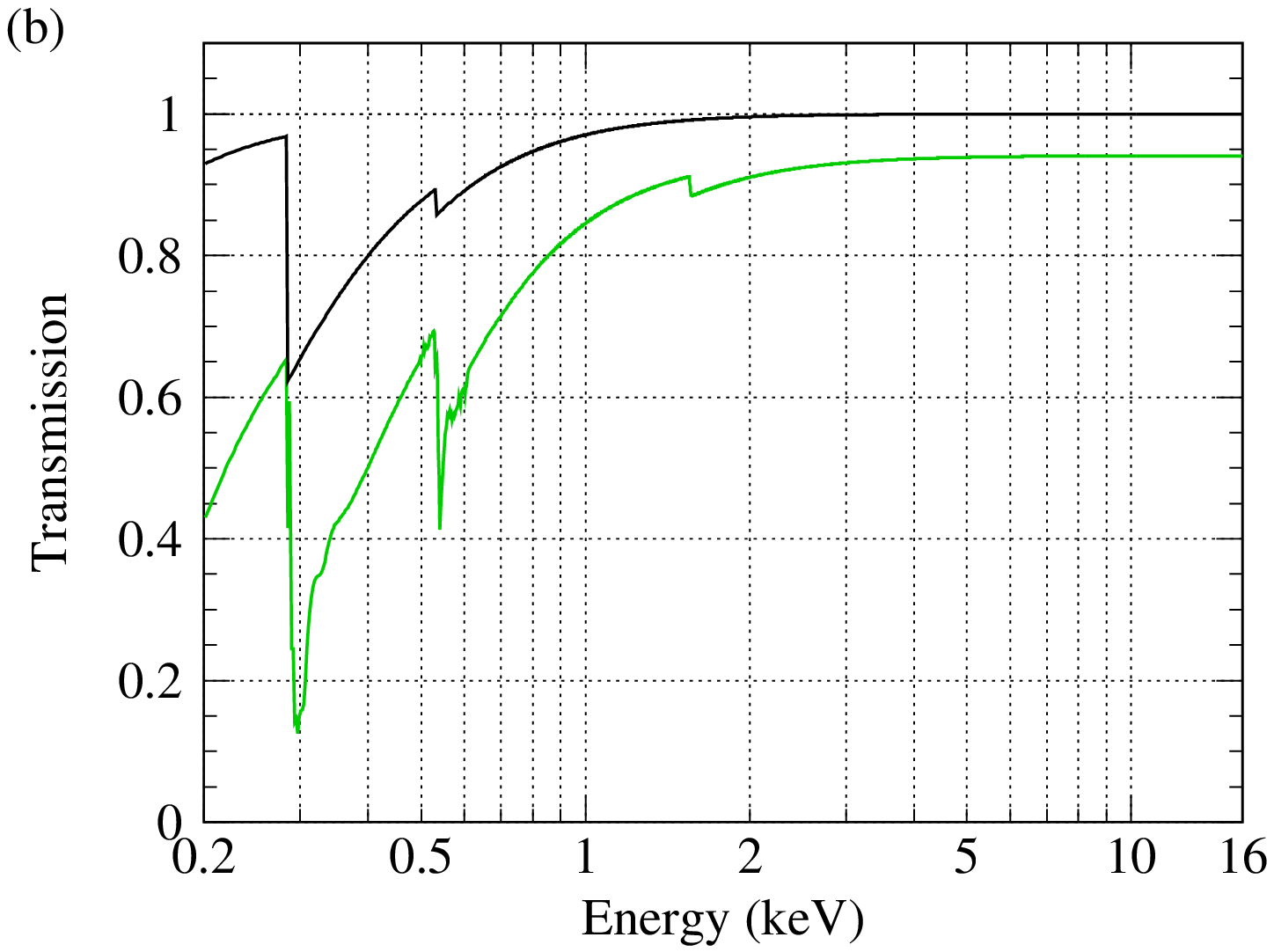}
}
\caption{
Plots of ARF columns generated for the Crab nebula,
with {\sf num\_photon}=100000 and {\sf estepfile\sc=sparse}
(55 energy steps) using the {\sl Chandra} image
in figure~\ref{fig:crab-image}~(a) as {\sf source\_image}.
The accumulation radius is 6~mm = 250 pixel $\simeq$ 4.34$'$
in the SKY coordinate.
(a) black: {\sc specresp}, green: {\sc xrt\_effarea},
cyan: 90\% confidence range of {\sc specresp}, namely,
$\makebox{\sc specresp}\pm 1.65\times \makebox{\sc resperr}$,
although it is almost hidden by the overlaid black line.
Red line indicates the nominal ARF in CALDB without contamination.
(b) green: {\sc shield\_transmis}, black: {\sc contami\_transmis}.
}\label{fig:crab-arf}
\end{figure*}

Using the {\sl Chandra} image, we also generated an ARF for XIS0,
and it is plotted in figures~\ref{fig:crab-arf}~(a) and (b).
The {\sc specresp} (black) and {\sc xrt\_effarea} (green) columns
are plotted in figure~\ref{fig:crab-arf}~(a), and
the {\sc contami\_transmis} (black) and {\sc shield\_transmis} (green)
columns are plotted in figure~\ref{fig:crab-arf}~(b).
The 90\% confidence range of the {\sc specresp} is also drawn
by cyan lines in figure~\ref{fig:crab-arf}~(a).
Full observational conditions, namely {\sf attitude},
{\sf gtifile}, {\sf aberration}, and {\sf contamifile},
are considered in the ARF generation with {\sf num\_photon}=100000
and {\sf estepfile\sc=sparse} (55 energy steps).
The accumulation radius is 6~mm = 250~pixel $\simeq$ 4.34$'$
in the SKY coordinate.

For comparison, we plot the nominal ARF in CALDB
without contamination, {\tt ae\_xi0\_xisnom6\_20060615.arf}, in red line.
In fact, the nominal ARF was also generated by an older version of
{\sl xissimarfgen}, and the calibration files were not changed
between the two versions. However, the nominal ARF is calculated
with much denser energy step (2~eV steps below 4~keV,
and at most 10~eV steps above 4~keV, with 3450 energy steps),
and 4 times higher photon statistics ({\sf num\_photon}=400000).
Although slight jerks are seen in black and green lines
in figure~\ref{fig:crab-arf}~(a), these two ARFs are quite consistent.
Discrepancy in the lower energy range is due to
the XIS contamination, which is plotted by a black line
in figure~\ref{fig:crab-arf}~(b).
Therefore, in the spectral fitting,
the Crab nebula can be treated as a point-like source with {\sl Suzaku} XIS,
if the extraction radius is large enough ($r\sim 6$~mm)
and the spacecraft attitude is stable.
We also note that the nominal ARFs give flux consistent
with that obtained by \citet{Toor1974} within $\sim 2$\%
for all the XIS sensors \citep{Serlemitsos2006}.

\begin{figure*}[p]
\centerline{
\FigureFile(0.9\textwidth,0.48\textwidth){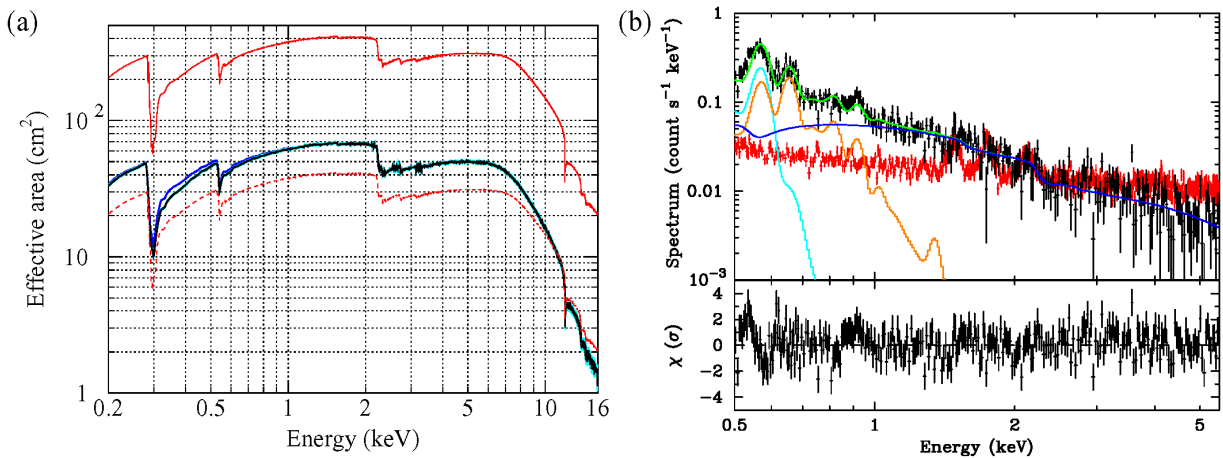}
}
\caption{
Plots of ARF columns generated for the NEP field,
with {\sf num\_photon}=400000, {\sf estepfile\sc=dense}
(2303 energy steps), {\sf source\_mode\sc=uniform},
and {\sc source\_rmax}=20$'$.
The accumulation region is all the XIS1 CCD including
the calibration source.
(a) black: {\sc specresp}, blue: {\sc specresp}\,/\,{\sc contami\_transmis},
cyan: 90\% confidence range of {\sc specresp}.
Solid red line indicates the nominal ARF in CALDB without contamination,
and dashed red line shows it multiplied by 0.1.
(b) Black line in the upper panel represent the NEP field spectrum
observed with {\sl Suzaku} in the ``stable'' period \citep{Fujimoto2006}.
The NXB is subtracted, and it is fitted by the
$[\; {\it apec}\makebox{ (cyan)} +
{\it apec}\makebox{ (orange)} +
{\it wabs}\times {\it power\makebox{-}law}\makebox{ (blue)} \;]$
model in {\sc Xspec} 11.3.2t indicated by green line.
The estimated NXB spectrum is overlaid in red line.
Fit residuals in units of $\sigma$ are shown in the lower panel.
}\label{fig:nep}
\end{figure*}

\subsection{NEP Field}\label{subsec:nep}

The NEP field is an archetypal ``blank field'',
where no X-ray bright objects exist. In such a region, the
X-ray background, including both the extra-galactic
\citep{Brandt2005} and the Galactic components
\citep{Snowden1995}, is the dominant X-ray source.
The X-ray background can be treated as almost uniform distribution,
hence we tested the uniform-sky ARF
generated by {\sl xissimarfgen} in this field.

We created an ARF assuming a uniform distribution for the source from a
circular region with a radius of $20'$ (see caption of
figure~\ref{fig:nep} for details of the parameters),
and fitted the observed spectrum with it.
In figure~\ref{fig:nep}~(a), the effective area, {\sc specresp},
of the obtained ARF is displayed in comparison with that for a point
source. One can see that the effective area is relatively smaller in the
higher energy band ($\gtrsim 7$~keV), due to the vignetting effect of
the XRT\@. After subtracting the NXB contribution
estimated using the night Earth database (\S\,\ref{subsec:notes-accum-region}),
the spectrum can be well fitted with a power-law model
representing the CXB and one or two thin-thermal plasma
models representing local Galactic thermal components,
as shown in figure~\ref{fig:nep}~(b).
The photon index and the surface brightness of the
power-law component are consistent with the parameters reported so far
\citep{Gendreau1995,Kushino2002}.
See Fujimoto et al. (2006) for the details of the analysis. This
result demonstrates that {\sl xissimarfgen} properly generates the ARF
for the uniform-sky emission.

\subsection{Abell~1060 Cluster of Galaxies}\label{subsec:a1060}

\begin{figure*}
\centerline{
\FigureFile(\textwidth,\textwidth){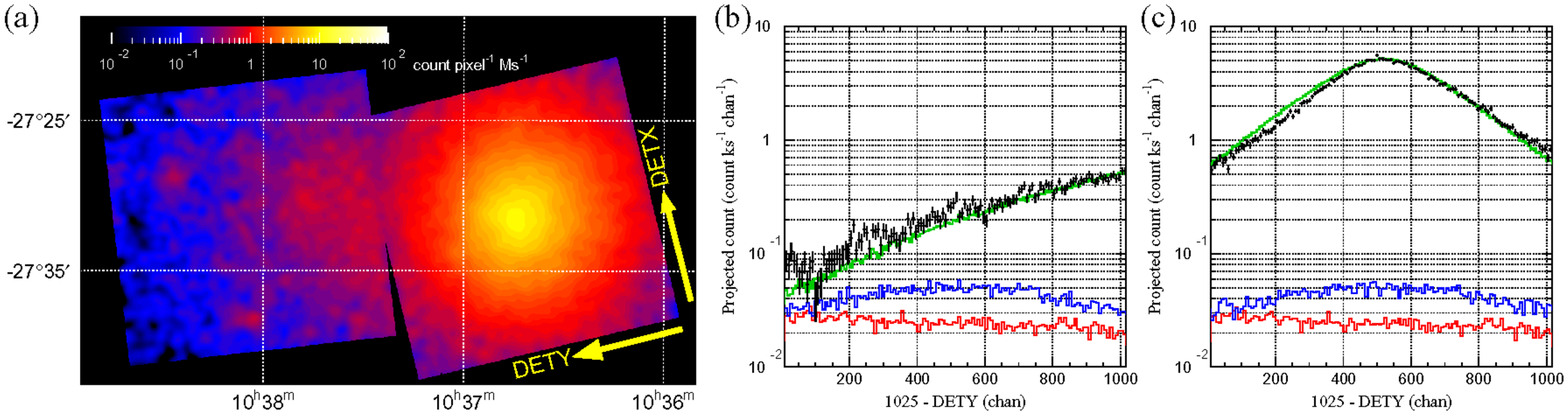}}
\caption{
(a) Observed Abell~1060 image combined for the central and offset pointings
obtained with XIS0 in the 1--4~keV energy range.
The image is smoothed with $\sigma=16$ pixel
$\simeq 17''$ Gaussian, and the estimated NXB and CXB components are
subtracted. The exposure time is corrected, but vignetting is not corrected.
Directions of DETX/Y axes are indicated in the figure.
(b) Comparison of the observed and the simulated images (1--4~keV)
projected to the DETY axis in the offset pointing.
The green line shows the simulated distribution by {\sl xissim}\/
assuming an analytical model (double-$\beta$ model) obtained with
{\sl XMM-Newton}, and the $kT=3.4$~keV {\it vapec} model spectrum.
The blue and red lines show the estimated CXB and NXB distribution,
respectively. The black crosses show the observed distribution
after subtracting the CXB and NXB components.
(c) Same as (b), but for the central observation.
}\label{fig:a1060-image}
\end{figure*}

Finally, we present an example of the Abell~1060 cluster of galaxies
observed with {\sl Suzaku}. Scientific results will be
published by K.~Sato et al.\ in preparation.
Abell~1060 is a circular and nearly isothermal ($\sim 3$~keV)
cluster of galaxies \citep{Tamura2000,Furusho2001,Hayakawa2004,Hayakawa2006}
and is suitable for testing the ARF for extended sources.
There were two pointings performed with {\sl Suzaku}
at the central region and the $\sim 20'$ east offset region,
as shown in figure~\ref{fig:a1060-image}~(a).
These observations were conducted at the end of November 2005,
when the XIS contamination was already significant
and was starting to saturate.

The observed spectrum is assumed to contain
{\sc(a)} thin thermal plasma emission from the intra cluster medium (ICM),
{\sc(b)} local Galactic emission, {\sc(c)} CXB, and {\sc(d)} NXB\@.
We can estimate {\sc(d)} using the night Earth database
mentioned in \S\,\ref{subsec:notes-accum-region},
and can subtract it from the observed spectrum.
As demonstrated in \S\,\ref{subsec:nep},
the spectrum of {\sc(b)} can be represented
by the ({\it apec} + {\it apec}) model
with 1~solar abundance, and that of {\sc(c)} has
a shape of absorbed power-law with $\Gamma\simeq 1.4$.
However, we cannot fit the observed spectrum directly
by a sum of {\sc(a)} + {\sc(b)} + {\sc(c)},
because the spatial distribution of these three are different on the sky,
as described in \S\,\ref{subsec:principle-and-limitations}.

We therefore adopted the following method for the spectral analysis
of Abell~1060. We extracted several spectra from annular regions
centered on the cluster core, and here we show two samples of
the innermost region at the projected radius of 0--2$'$
from the central observation,
and the outermost region of 17--27$'$
from the offset observation,
as representatives.
We generated two different ARFs for each spectrum,
$A^{\makebox{\small\sc u}}(E_i)$ and
$A^{\makebox{\small\sc b}}(E_i)$,
which respectively assume the uniform-sky emission and
the ICM surface brightness profile obeying
an analytical model obtained with the {\sl XMM-Newton} data.

As described in \S\,\ref{subsec:principle-and-limitations},
it is important that the assumed spatial distribution on the sky
well agrees with the actual data in the calculation of the ARF response.
We therefore compared the observed images with the simulated ones
in figure~\ref{fig:a1060-image}~(b) and (c).
The 1--4~keV energy range was chosen so that the distortion of the
image due to the XIS contamination and the XRT vignetting was not severe.
In this energy range, the Galactic component {\sc(b)} is almost
negligible, whereas the CXB and NXB components cannot be neglected
especially in the offset observation.
The NXB component (red line) is estimated from the night Earth database.
A small ACTY (= DETY\,$- 1$ for XIS0, see figure~\ref{fig:raw/act/det})
dependence of the NXB intensity is seen, reflecting the dwell time
at the frame-store region of the CCD\@.
The CXB component (blue line) is estimated by the {\sl xissim}\/
simulation, assuming the uniform sky and the previous {\sl ASCA} results
of the CXB intensity \citep{Kushino2002}.
The vignetting effect is seen in the CXB counts,
hence the count rate slightly drops at the CCD rim.
After the subtraction of the estimated CXB and NXB components,
the observed distribution of the cluster (black crosses) is
fairly well reproduced by the {\sl xissim}\/ simulation of the
cluster emission (green line), although a small asymmetry is observed
for the real cluster in the central observation.

\begin{figure*}
\centerline{
\FigureFile(\textwidth,\textwidth){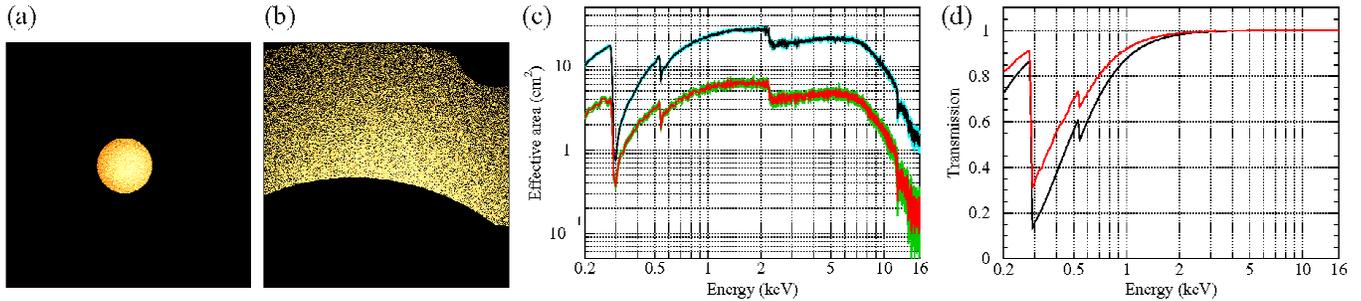}}
\caption{
Plots of the XIS1 ARFs for the Abell~1060 cluster of galaxies
calculated with {\sf limit\_mode\sc=mixed}, {\sf num\_photon}=100000,
{\sf accuracy=0.005}, and {\sf estepfile\sc=dense}.
(a)~The primary extension image in DET coordinate ($1024\times 1024$)
for the central observation at the projected radius of $r < 2'$.
(b)~Same as (a) but
for the offset observation at the projected radius of $17'<r<27'$.
(c)~The {\sc specresp} columns for
the central (black) and offset (red) observations plotted against energy.
The 90\% confidence range for each ARF is indicated by cyan or green
lines, respectively.
(d)~The {\sc contami\_transmis} columns for
the central (black) and offset (red) observations.
}\label{fig:a1060-arf}
\end{figure*}

Figures~\ref{fig:a1060-arf}~(a)--(d) show
the latter kind of ARFs, $A^{\makebox{\small\sc b}}(E_i)$,
for both regions.
Figures~\ref{fig:a1060-arf}~(a) and (b) correspond to
the extraction regions of 0--2$'$ and 17--27$'$, respectively,
in the DET coordinate, and the calibration source area
(top-left and bottom-right for XIS1) is also excluded in (b).
Although the accumulation area is smaller for (a) than (b),
the calculated effective area is much larger for (a) than (b)
as seen in figure~\ref{fig:a1060-arf}~(c),
plotted in black and red lines, respectively,
due to the assumed surface brightness profile.
One can see the position dependence of the XIS contamination
(thinner towards the CCD edge) is treated appropriately
as seen in figure~\ref{fig:a1060-arf}~(d).

Denoting the spectra of {\sc(a)}, {\sc(b)}, {\sc(c)}, and {\sc(d)} as
$M^{\makebox{\tiny\sc icm}}(E_i)$,
$M^{\makebox{\tiny\sc gal}}(E_i)$,
$M^{\makebox{\tiny\sc cxb}}(E_i)$, and
$\mathcal{M}^{\makebox{\tiny\sc nxb}}({\it PI}_j)$,
the observed spectrum can be expressed by a sum of,
\begin{eqnarray}
A^{\makebox{\small\sc b}}\otimes M^{\makebox{\tiny\sc icm}} +
A^{\makebox{\small\sc u}}\otimes M^{\makebox{\tiny\sc gal}} + 
A^{\makebox{\small\sc u}}\otimes M^{\makebox{\tiny\sc cxb}} +
\mathcal{M}^{\makebox{\tiny\sc nxb}},
\end{eqnarray}
where the operator $\otimes$ denotes the transformation
defined by eq.~(\ref{eq:rmf-operation}).
It is known that the CXB spectrum, $M^{\makebox{\tiny\sc cxb}}$,
is fairly constant over the sky except for difference in
the neutral hydrogen column density, $N_{\rm H}$, for absorption,
whereas the local Galactic emission, $M^{\makebox{\tiny\sc gal}}$,
may vary from field to field by more than an order of magnitude
\citep{Kushino2002}.

\begin{figure*}
\centerline{
\FigureFile(\textwidth,\textwidth){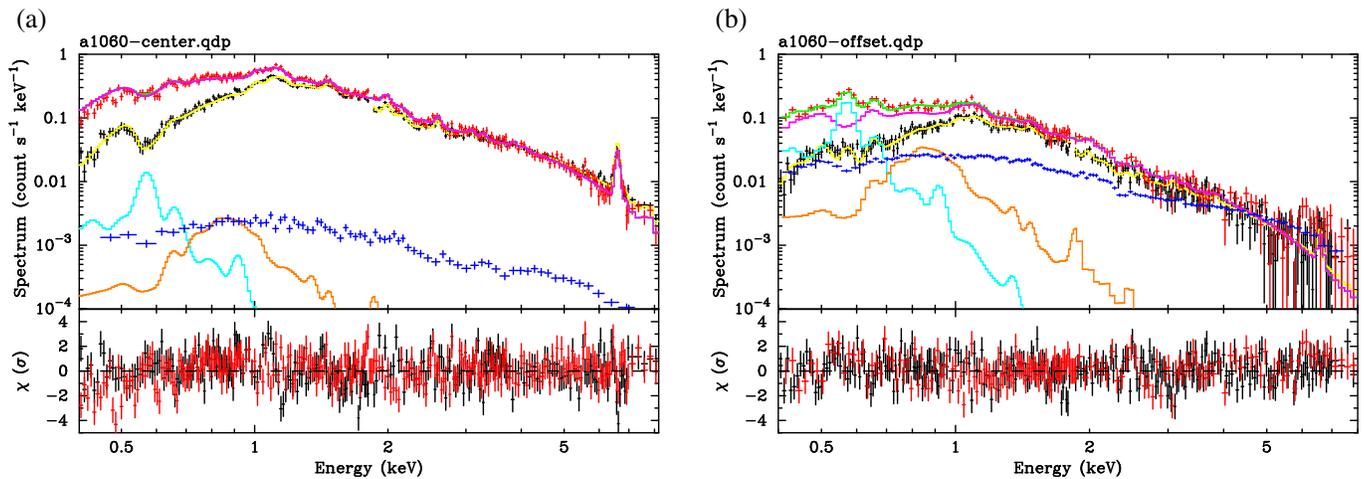}}
\caption{
Example spectra of the Abell~1060 cluster of galaxies,
(a) for the central observation, (b) for the offset observation.
In both figures, red or black crosses represent the observed
spectrum with the XIS1 (BI) or XIS0+XIS2+XIS3 (FI) sensor(s), respectively,
for the upper panels, and the fit residuals for the lower panels.
The estimated CXB + NXB spectrum has been subtracted from
each observed spectrum, and the estimated CXB spectra for XIS1
are indicated by blue crosses. The spectra are fitted by the
$[\; {\it apec}\makebox{ (cyan)} +
{\it apec}\makebox{ (orange)} +
{\it phabs}\times {\it vapec}\makebox{ (magenta)} \;]$
model in {\sc Xspec} 11.3.2t indicated by green line for XIS1
and yellow line for the FI sensors.
The model components are only plotted for the XIS1 spectrum.
}\label{fig:a1060-spec}
\end{figure*}

Considering this situation, we assumed a power-law spectrum
for the CXB with the values by \citet{Kushino2002},
$\Gamma=1.4$ and $S_{\rm X} = 5.97\times 10^{-8}$
erg~cm$^{-2}$~s$^{-1}$~sr$^{-1}$ (2--10~keV),\footnote{
This value is taken from table 3 of \citet{Kushino2002},
for the integrated spectrum with source elimination
brighter than $S_0=2\times 10^{-13}$ erg~cm$^{-2}$~s$^{-1}$ (2--10~keV)
in the GIS filed of view with $\Gamma=1.4$ (fix)
and the nominal NXB level (0\%).}
measured with the {\sl ASCA} GIS \citep{Ohashi1996,Makishima1996}.
The neutral hydrogen column density was fixed to
$N_{\rm H}= 4.9\times 10^{20}$ cm$^{-2}$ \citep{DickeyLockman1990}.
We calculated the estimated contribution of the CXB,
$A^{\makebox{\small\sc u}}\otimes M^{\makebox{\tiny\sc cxb}}$,
using the {\sl fake} command in {\sc Xspec}.
This contribution for each region is indicated by blue crosses
in figures~\ref{fig:a1060-spec}~(a) and (b) for XIS1.
We subtracted the CXB contribution from the observed spectrum
as well as the estimated NXB spectrum.
The XIS1 (red) and FI (XIS0+XIS2+XIS3; black) spectra
in figures~\ref{fig:a1060-spec}~(a) and (b) denote
those after the CXB and NXB subtraction.
We then fitted the XIS1 and FI spectra simultaneously
for the offset observation, where the Galactic component
{\sc(b)} is prominent, with the $[\; {\it apec}\makebox{ (cyan)} +
{\it apec}\makebox{ (orange)} +
{\it phabs}\times {\it vapec}\makebox{ (magenta)} \;]$ model,
using the ARF response, $A^{\makebox{\small\sc b}}(E_i)$.

In a strict sense, using $A^{\makebox{\small\sc b}}$
for the component {\sc(b)} is not correct,
which should be $A^{\makebox{\small\sc u}}$ instead.
We made this choice due to the limitation of
the {\sc Xspec} v11 (\S\,\ref{subsec:principle-and-limitations}),
however, it does not matter practically if we only
notice the shape of the spectrum.
The absolute surface brightness of the Galactic component
was evaluated separately using the {\sc Xspec} {\sl fake}\/ command
and the $A^{\makebox{\small\sc u}}$ response.
We then fitted the central region, fixing the shape of the
Galactic component, but with its normalization scaled so that
the surface brightness is preserved between the two different sky regions.
{\sc Xspec} v12 can handle this situation more straightforwardly.

The released RMF, {\tt ae\_xi[0-3]\_20060213.rmf},
was used for the spectral analysis.
The ARFs were generated by {\sl xissimarfgen},
and were convolved with the RMFs and added for three FI sensors,
using the {\sl marfrmf} and {\sl addrmf} tasks in {\sc Ftools}.
As for the photon statistics of the simulation,
{\sf limit\_mode={\sc mixed}} was chosen
with {\sf num\_photon}=100000 and {\sf accuracy}=0.005.
As seen in figures~\ref{fig:a1060-spec}~(a) and (b),
both the observed spectra can be well fitted by
one temperature plasma emission model for ICM,
and ({\it apec} + {\it apec}) model with 1~solar abundance
for the local Galactic emission.
The surface brightness and the spectral shape of the Galactic emission
is kept constant between both regions.
This is confirmed by the fact the ratio of the Galactic components
to the CXB is almost equal between figures~\ref{fig:a1060-spec}~(a) and (b).
Note that the fit appears equally good to the BI (XIS1) and FI sensors,
which are different in the thickness of the contamination.
So far, it has been confirmed that the temperature for each ring derived from
the spectral fitting with this method is quite consistent with
the previous results with {\sl XMM-Newton} and {\sl Chandra}
\citep{Hayakawa2004,Hayakawa2006}.
See K.~Sato et al.\ in preparation for details of the results.

\section{Summary}\label{sec:summary}

\begin{itemize}
\item
We have developed a Monte-Carlo simulator of the {\sl Suzaku} XRT/XIS system
taking into account full calibration results.
\\
\item
We adopted the ANL platform that provides us a flexible and comprehensive 
environment for the {\sl Suzaku} software production. 
\\
\item
There is a dedicated task named {\sl mkphlist}
which generates a photon file to feed to the simulator.
\\
\item
The task {\sl xissim} reads the photon file, and conducts the instrumental
simulation using the XRT ray-tracing library and
the RMF of the XIS, and generates an event file,
which is consistent with that for real observation,
so that users can analyze the simulated data
in the same manner as real data.
\\
\item
The simulator-based ARF generator is named {\sl xissimarfgen},
which can compute up to 200 ARFs corresponding
to different accumulation regions by a single batch of simulations.
\\
\item
The combination of {\sl xissim} and {\sl xissimarfgen}
enables users to analyze spatially extended and spectroscopically
complex celestial sources. 
\\
\item
Since one of the {\sl Suzaku}'s unique features is
the low and stable particle background, 
these simulators are crucial for producing scientific results
with low signal-to-noise data from extended sources. 
\\
\item
The latest public version is 2006-08-26, which will be included in
the next official release of the {\sl Suzaku} {\sc Ftools}
scheduled in late 2006.
\end{itemize}

\bigskip

Thanks are given to the referee,
Dr.~K.~Arnaud, for useful comments which improved the original manuscript.
We express sincerely thanks to H.~Honda
for early stage work on the ASCA\_ANL, and SimASCA\@.
We also show R.~L.~Fink our appreciation for early stage work
on the {\sl xrrt} ray-tracing library.
We acknowledge to L.~Angelini and I.~Harrus for useful discussion
on the CALDB file format.
We thank K.~Mori to provide us a Chandra image of the Crab nebula
corrected for pile-up photons.
We also acknowledge D.~McCammon, R.~Smith, Y.~Takei, C.~Matsumoto,
and N.~Ota for testing and giving valuable comments
on the {\sl xissim} and {\sl xissimarfgen} tasks.
Part of this work was financially supported by the Ministry of
Education, Culture, Sports, Science and Technology of Japan,
Grant-in-Aid for Scientific Research No.\ 14079103, 15001002.

\appendix
\section{Definition of the Coordinates}\label{app:coordinates}

\begin{figure*}
\centerline{\FigureFile(0.8\textwidth,\textwidth){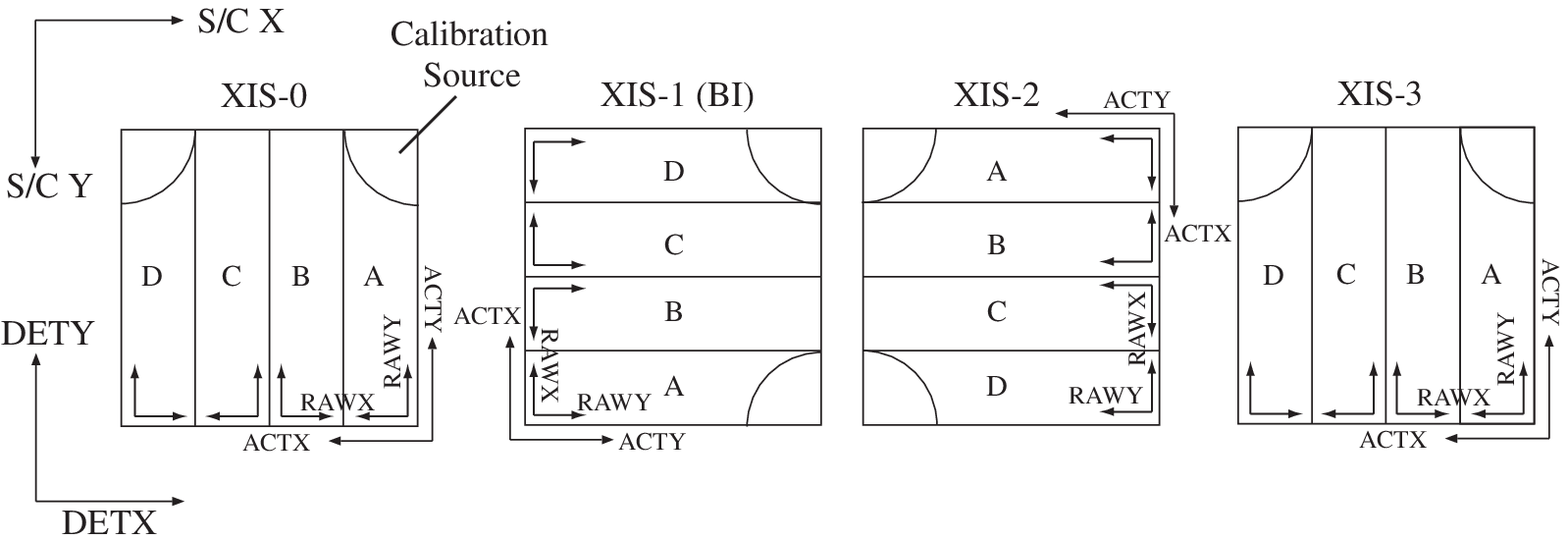}}
\caption{
Relations between RAWX/Y, ACTX/Y, DETX/Y among the four XIS sensors.
The coordinate are defined looking up from the XIS toward the XRT\@.
}\label{fig:raw/act/det}
\end{figure*}

\begin{table}
\caption{
Summary of XIS coordinate column information.
}\label{tab:coord-column}
\centerline{\small
\begin{tabular}{lcccc}
\hline\hline
\makebox[5.5em][c]{Column Name}
		&Min\makebox[0in][l]{$\;^*$}
				& Max\makebox[0in][l]{$\;^\dagger$}
					& Origin\makebox[0in][l]{$\;^\ddagger$}
							& Pixel Size$\;^\S$ \\
\hline
{\sc segment}	&0		& 3	 	& --	& --\\
{\sc rawx/y}	&0		& 255/1023     	& --	& 0.024 mm\\
{\sc actx/y}	&0		& 1023   	& --	& 0.024 mm\\
{\sc detx/y}	&1		& 1024    	& 512.5	& 0.024 mm\\
{\sc focx/y}	&1		& 1536		& 768.5	& 0.024 mm\\
{\sc x/y}	&1\makebox[0in][l]{$\;^\|$}
		&1536\makebox[0in][l]{$\;^\|$}
		& 768.5	& 0.0002895 deg\makebox[0in][l]{$\;^\sharp$}\\
\hline\\[-1ex]
\multicolumn{5}{l}{\parbox{0.47\textwidth}{\footnotesize
\footnotemark[$*$] TLMIN$n$ keywords in the event file.}}\\
\multicolumn{5}{l}{\parbox{0.47\textwidth}{\footnotesize
\footnotemark[$\dagger$] TLMAX$n$ keywords in the event file.}}\\
\multicolumn{5}{l}{\parbox{0.47\textwidth}{\footnotesize
\footnotemark[$\ddagger$] TCRPX$n$ keywords in the event file.}}\\
\multicolumn{5}{l}{\parbox{0.47\textwidth}{\footnotesize
\footnotemark[$\S$] TCDLT$n$ keywords in the event file.}}\\
\multicolumn{5}{l}{\parbox{0.47\textwidth}{\footnotesize
\footnotemark[$\|$] Default image region.
X/Y values can be outside of the region.}}\\
\multicolumn{5}{l}{\parbox{0.47\textwidth}{\footnotesize
\footnotemark[$\sharp$] Angular scale at the center.
Outer pixels are slightly different due to the tangential projection.}}
\end{tabular}
}
%\end{table}
\bigskip
%\begin{table}
\caption{
Summary of XIS alignment information
}\label{tab:teldef}
\centerline{\small
\begin{tabular}{ll}
\hline\hline
Item				& Ideal Value \\
\hline
Focal length			&  4750 mm \\
Optical axis location in DET	& (512.5, 512.5) \\
Size of the DET pixel		&  0.024 mm/pixel \\
Offsets between DET and FOC	& (0.0, 0.0)\\
Roll angle between DET and FOC	&  0.0 deg \\
Alignment matrix for FOC $\rightarrow$ SKY\makebox[0in][l]{$\;^*$}
				& $3\times3$ identity matrix \\
\hline\\[-1ex]
\multicolumn{2}{l}{\parbox{0.47\textwidth}{\footnotesize
\footnotemark[$*$] Alignment matrix is common to all sensors.}}
\end{tabular}
}
\end{table}

The following coordinates are defined to describe event locations in
the telemetry, on the detector, or on the sky.

\smallskip\noindent{\bf RAW coordinates:}
Original digitized values in the telemetry to identify the pixels of the
events.  This may not reflect physical locations of the pixels on the
sensor.  For example, XIS RAWX (or RAWY) coordinate will have values
from 0 to 255 (or 1023) on each CCD segment.
Each of the four XIS sensors has a single CCD chip,
and a single chip is divided into four segments\@.

\smallskip\noindent{\bf ACT coordinates:}
The ACTX/Y values are defined to represent actual pixel
locations in the CCD chips.
ACTX/Y will take 0 to 1023  to denote the 1024 $\times$ 1024
pixels in the chip.  The XIS RAW to ACT conversion depends on the
observation modes (such as Window Options) and will require
housekeeping information.
The XIS ACT coordinate is defined by looking down on the sensors, hence 
the ACTX/Y to DETX/Y conversion needs a flip in the Y-direction.

\smallskip\noindent{\bf DET coordinates:}
Physical positions of the pixels within each sensor, XIS0--3\@.
Misalignments between the sensors are not taken into account.
The DETX/Y coordinate are defined by looking up the sensor, such that
the spacecraft (S/C) +Y direction becomes the $-$DETY direction
(the same convention as with ASCA ). The S/C Z-axis points in the
telescope direction, and +Y direction is toward the solar paddle.
For XIS, the DETX and DETY values take 1 to 1024.

\smallskip\noindent{\bf FOC coordinates:}
Focal plane coordinate common to all the sensors. Misalignments
between the sensors are taken into account so that the FOC images
of different sensors can be superposed.
The origin of the FOC coordinate corresponds to the XIS nominal
position for pointing observations.
FOC is calculated from DET by linear transformation to represent
the instrumental misalignment, i.e., the offset and the roll angle.

\smallskip\noindent{\bf SKY coordinate:}
Positions of the events on the sky.
For each XIS event, the equatorial coordinate of the pixel center
projected on a tangential plane are given.
The aberration correction due to parallax (i.e., the revolution
of the Earth around the Sun) is also considered.

\smallskip\noindent{\bf XRT coordinate:}
This is given by (XRTX, XRTY) in mm on the focal plane,
or ($\theta$, $\phi$) corresponding to the offset angle ($'$)
and the azimuth angle ($^\circ$)
with respect to the optical axis of each XRT\@.
The location of the optical axis on the DET coordinate is defined
so that effective area of the XRT is maximized.

\smallskip
The RAW, ACT, DET, FOC and SKY coordinate are written in
the {\sl Suzaku} XIS event files.
Relations between RAWX/Y, ACTX/Y, DETX/Y among the four XIS sensors
are summarized in figure~\ref{fig:raw/act/det}.
The  DETX/Y  pixel sizes correspond to the physical pixel size
of the XIS CCD, while the X/Y pixel size corresponds to the angular
scale of a single CCD pixel at the reference pixel.
To allow rotation of the image and some shift
of the pointing direction during the observation,
the X/Y range is taken slightly bigger than $\sqrt{2}\times 1024$.
The minimum value, maximum value, origin of the coordinate
(reference pixel location), and pixel size are summarized in
table~\ref{tab:coord-column}.

There is a file called {\it teldef}\/ (namely, telescope definition)
for each sensor.  In the primary header of each {\it teldef}\/ file,
alignment data for the individual sensors
(DET$\rightarrow$FOC, FOC$\rightarrow$SKY, and DET$\rightarrow$XRT)
are given. The alignment parameters in the {\it teldef}\/ file
are summarized in table~\ref{tab:teldef}.
In the extensions of the {\it teldef}\/ files, sensor-dependent additional
calibration information may be written. 
For example, the 1st extension of the XRS {\it teldef}\/ file
has measured positions and sizes of the XRS pixels.

In this scheme, the conversion from RAW to
DET does not depend on the misalignments between the sensors.
Therefore, DETX/Y, as well as RAWX/Y, can be written in the event
files without having the calibration information.
The DET to FOC conversion requires the sensor misalignment
data. 
The conversion from FOC to SKY is made using the satellite
Z-Y-Z Euler angles (${\it ea}_1$, ${\it ea}_2$, ${\it ea}_3$)
in the attitude file and the 3$\times$3 alignment matrix given
in the {\it teldef}\/ file.
One must be careful because this conversion is dependent on
the observation date and direction due to the parallax (aberration) correction.
The magnitude of the correction is about $\pm 20.5''$ at maximum.

All the conversions between these coordinates are supplied in
the form of the C functions in the {\sl astetool}\/ library.
These functions make use of the information given by the {\it teldef}\/ file,
and it is strongly recommended to use them for the coordinate conversions.
They are built on the ISAS-made mission-independent library named
{\sl atFunctions}, which includes basic routines to
handle 3-dimensional vectors and rotation matrices.
There is also a frontend of the coordinate
conversions in the {\sl Suzaku} {\sc Ftools}, named {\sl aecoordcalc}.

\section{Structures \& Parameters}\label{app:structure}

\subsection{mkphlist}
\label{subsec:mkphlist-structure}

The {\sl mkphlist} task consists of three ANL modules as
listed in table~\ref{tab:mkphlist}.
The SimASTE\_Root (we will omit SimASTE\_ hereafter in the main text)
module is a root module for the {\sl Suzaku} simulators,
that handles initialization of random numbers and common CALDB files.
The PhotonGen module generates photons according to
the parameters set by a user, and caches the photon parameters
in an internal storage area called BNK \citep{Ozaki2006}\@.
The PhotonFitsWrite module retrieves the photon data from the BNK
and writes the data to the photon file.
By splitting these functions into dedicated ANL modules,
it is easier to understand the structure of the task,
and furthermore we can share the modules among several tasks.
For example, the Root module is used for all the SimASTE tasks,
and the PhotonGen modules is shared with {\sl xissim}.

The parameters of the PhotonGen module (table~\ref{tab:mkphlist})
is classified into
the following five groups:
{\it (1)} to determine the X-ray flux,
{\sf photon\_flux}, {\sf flux\_emin}, {\sf flux\_emax},
and {\sf geometrical\_area};
{\it (2)} to determine the spectral shape of incident X-rays,
{\sf spec\_mode}, {\sf qdp\_spec\_file}, and {\sf energy};
{\it (3)} to determine the spatial distribution on the sky,
{\sf image\_mode}, {\sf ra}, {\sf dec},
{\sf sky\_r\_min}, {\sf sky\_r\_max},
{\sf fits\_image\_file};
{\it (4)} to determine the photon arrival time to be
equal or random interval steps,
{\sf time\_mode};
{\it (5)} to determine how many photons are to be generated,
{\sf limit\_mode}, {\sf nphoton}, and {\sf exposure}.

\subsection{xissim}
\label{subsec:xissim-structure}

Table~\ref{tab:xissim} summarizes the ANL modules and major parameters
for {\sl xissim}. It consist of eight modules, the first two modules
of which are common to {\sl mkphlist}.

In the Root module, the {\sf simulation\_mode}, {\sf instrume},
{\sf teldef}, and {\sf leapfile} parameters are added
(which are ignored in {\sl mkphlist})
when compared with table~\ref{tab:mkphlist}.
The {\sf simulation\_mode} parameter determines the default
mode of the simulation, and the two defined modes are 
{\sc discard} and {\sc weight}. In the {\sc discard} mode,
each absorbed photon is discarded, for example,
by absorption in the XRT thermal shield.
In contrast, the {\sc weight} of the photon is decreased
by multiplying the transmission probability of the thermal shield
in the {\sc weight} mode.
The final value of the {\sc weight} is written to the {\sc weight} column
of the output event file.
This feature enables efficient simulation
when most of photons disappear during the simulation,
however one needs to use care in the handling of the simulation results.
The default {\sf simulation\_mode} is {\sc discard} for {\sl xissim},
whereas {\sf simulation\_mode\sc=weight} for {\sl xissimarfgen}
to treat the thermal shield transmission separately
(\S\,\ref{subsec:xissimarfgen-implement}).

The PhotonGen module enables on-the-fly photon generation without
input photon files, and is usually deactivated ({\sf enable\_photongen}=no). 
The same parameters in table~\ref{tab:mkphlist} are usable in this mode.
The PhotonRead module reads up to eight photon files,
as well as the GTI file and the attitude file,
and puts the photon data ({\sc ra}, {\sc dec},
{\sc photon\_time}, {\sc photon\_energy})
and the Euler angles at {\sc photon\_time} into BNK\@.
By mixing multiple photon files, it is capable of simulating an
observation, e.g.\ hot and widely extended emission from a cluster of
galaxies with cool emission from the core region.

The ECStoXRTIN module takes care of the pre-XRT component.  It retrieves the
photon data and the Euler angles, 
and converts the photon positions into ($\theta$, $\phi$).
The parallax (aberration) correction and the $\cos\theta$ effect
are also considered here.
XRTsim conducts the ray-tracing by calling the {\sl xrrt} library,
and the XRTOUTtoDET compute the detector position hit by the photon.
XISRMFsim simulates the XIS using the RMF,
and XISevtFitsWrite write the final output (table~\ref{tab:event-file})
into the event file.

\begin{table}
\caption{
Structure and parameters of {\sl mkphlist}.
}\label{tab:mkphlist}
\centerline{\small
\begin{tabular}{lll}\hline\hline
\multicolumn{2}{l}{\makebox[0in][l]{%
Module/{\sf Parameter}\makebox[0in][l]{$\;^*$}}} &
				Description \\
\hline
\makebox[0in][l]{SimASTE\_Root} & \\
& ({\sf rand\_seed}) &		random number seed \\
& ({\sf rand\_skip}) &		random number skip count \\
\makebox[0in][l]{SimASTE\_PhotonGen} & \\
& {\sf photon\_flux} &		photon flux (photons~cm$^{-2}$~s$^{-1}$) \\
& {\sf flux\_emin} &		lower energy (keV) for {\sf photon\_flux} \\
& {\sf flux\_emax} &		upper energy (keV) for {\sf photon\_flux} \\
& {\sf geometrical\_area}\hspace*{-0.5em} & XRT geometrical area (cm$^2$) \\
& {\sf spec\_mode} &		0:qdp-spec, 1:monochrome \\
& {\sf qdp\_spec\_file} &	qdp spectral file for {\sf spec\_mode}=0 \\
& {\sf energy} &		energy (keV) for {\sf spec\_mode}=1 \\
& {\sf image\_mode} &		0:FITS-image, 1:point, 2:uniform \\
& {\sf ra, dec} & {\sc ra}, {\sc dec} ($^\circ$) for {\sf image\_mode}=1 or 2\\
& {\sf sky\_r\_min} &		min radius ($'$) for {\sf image\_mode}=2 \\
& {\sf sky\_r\_max} &		max radius ($'$) for {\sf image\_mode}=2 \\
& {\sf fits\_image\_file} &	image FITS file for {\sf image\_mode}=1 \\
& {\sf time\_mode} &		0:constant, 1:Poisson \\
& {\sf limit\_mode} &		0:number of photon, 1:exposure time \\
& {\sf nphoton} &		number of photon for {\sf limit\_mode}=0 \\
& {\sf exposure} &		exposure time (s) for {\sf limit\_mode}=1 \\
\makebox[0in][l]{SimASTE\_PhotonFitsWrite} & \\
& {\sf outfile} &		output photon file name \\
\hline\\[-1ex]
\multicolumn{3}{l}{\parbox{0.47\textwidth}{\footnotesize
\footnotemark[$*$] Parameters in parentheses are hidden parameters.}}
\end{tabular}
}
\end{table}

\begin{table}
\caption{
Structure and parameters of {\sl xissim}.
}\label{tab:xissim}
\centerline{\small
\begin{tabular}{lll}\hline\hline
\multicolumn{2}{l}{\makebox[0in][l]{%
Module/{\sf Parameter}\makebox[0in][l]{$\;^*$}}} &
				Description \\
\hline
\makebox[0in][l]{SimASTE\_Root} \\
& {\sf instrume} &		instrument ({\sc xis0,xis1,xis2,xis3}) \\
& ({\sf simulation\_mode})\hspace*{-0.5em} &
			0:{\sc discard}, 1:{\sc weight} \\
& ({\sf rand\_seed}) &		random number seed \\
& ({\sf rand\_skip}) &		random number skip count \\
& ({\sf teldef}) &		{\it teldef}\/ file name \\
& ({\sf leapfile}) &		leap second file \\
\makebox[0in][l]{SimASTE\_PhotonGen $^\dagger$} \\
& ({\sf enable\_photongen})\hspace*{-1em} & enable on-the-fly photon generation \\
\makebox[0in][l]{SimASTE\_PhotonRead} \\
& {\sf infile$N$} &		input photon file(s) up to $N=8$ \\
& ({\sf gtifile}) &		name of the GTI file or {\sc none} \\
& ({\sf date\_obs}) &	observation start for {\sf gtifile}={\sc none}\\
& ({\sf date\_end}) &	observation end for {\sf gtifile}={\sc none}\\
& ({\sf attitude}) &		name of the attitude file or {\sc none} \\
& {\sf ea1}, {\sf ea2}, {\sf ea3} &	Euler angles for {\sf attitude}={\sc none} \\
& ({\sf pointing}) &		pointing type, {\sc auto} or {\sc user} \\
& {\sf ref\_alpha} &	{\it skyref}\/ {\sc ra} ($^\circ$) for {\sf pointing}={\sc user} \\
& {\sf ref\_delta} &	{\it skyref}\/ {\sc dec} ($^\circ$) for {\sf pointing}={\sc user} \\
& ({\sf ref\_roll}) &	{\it skyref}\/ {\sc roll} ($^\circ$) for {\sf pointing}={\sc user}\\
\makebox[0in][l]{SimASTE\_ECStoXRTIN} \\
& ({\sf aperture\_cosine}) &	consider aperture decrease by $\cos\theta$ \\
& ({\sf aberration}) $^\ddagger$ &	enable the aberration correction \\
\makebox[0in][l]{SimASTE\_XRTsim} \\
& ({\sf shieldfile}) &		XRT thermal shield transmission file \\
& ({\sf mirrorfile}) &		XRT mirror geometry file \\
& ({\sf reflectfile}) &		XRT surface reflectivity file \\
& ({\sf backproffile}) &	XRT backside scatter profile file \\
\makebox[0in][l]{SimASTE\_XRTOUTtoDET} \\
\makebox[0in][l]{SimASTE\_XISRMFsim} \\
& {\sf xis\_rmffile} &		XIS RMF name \\
& ({\sf aberration}) $^\ddagger$ &	enable the aberration correction \\
& ({\sf xis\_contamifile}) &	XIS contamination file or {\sc none}\\
& ({\sf xis\_efficiency}) &	multiply XIS effciency or not\\
& ({\sf xis\_chip\_select}) &	discard events fallen outside of CCD\\
\makebox[0in][l]{SimASTE\_XISevtFitsWrite} \\
& {\sf outfile} &		output event file name \\
\hline\\[-1ex]
\multicolumn{3}{l}{\parbox{0.47\textwidth}{\footnotesize
\footnotemark[$*$]
Parameters in parentheses are hidden parameters.}}\\
\multicolumn{3}{l}{\parbox{0.47\textwidth}{\footnotesize
\footnotemark[$\dagger$]
See table~\ref{tab:mkphlist} for rest of parameters when
{\sf enable\_photongen}=yes.}}\\
\multicolumn{3}{l}{\parbox{0.47\textwidth}{\footnotesize
\footnotemark[$\ddagger$]
The {\sf aberration} parameter is read in two modules.}}
\end{tabular}
}
\end{table}

\begin{table}
\caption{
Structure and parameters of {\sl xissimarfgen}.
}\label{tab:xissimarfgen}
\centerline{\small
\begin{tabular}{lll}\hline\hline
\multicolumn{2}{l}{\makebox[0in][l]{%
Module/{\sf Parameter}}} &	Description \\
\hline
\makebox[0in][l]{SimASTE\_Root $^\dagger$} \\
\makebox[0in][l]{SimASTE\_XISarfPhotonGen} \\
& ({\sf pointing}) &	pointing type, {\sc auto} or {\sc user} \\
& {\sf ref\_alpha} & {\it skyref}\/ {\sc ra} ($^\circ$) for {\sf pointing\sc=user}\\
& {\sf ref\_delta} & {\it skyref}\/ {\sc dec} ($^\circ$) for {\sf pointing\sc=user}\\
& ({\sf ref\_roll})& {\it skyref}\/ {\sc roll} ($^\circ$) for {\sf pointing\sc=user}\\
& {\sf source\_mode} &	{\sc skyfits,detfits,} \\
& &			{\sc j2000,skyxy,detxy,uniform} \\
& {\sf source\_image} &	FITS image for {\sf source\_mode}={\sc *fits}\\
& {\sf source\_ra} &	{\sc ra} ($^\circ$) for {\sf source\_mode\sc=j2000}\\
& {\sf source\_dec} &	{\sc dec} ($^\circ$) for {\sf source\_mode\sc=j2000}\\
& {\sf source\_x} &	$x$ (pixel) for {\sf source\_mode\sc=*xy}\\
& {\sf source\_y} &	$y$ (pixel) for {\sf source\_mode\sc=*xy}\\
& {\sf source\_rmin} &	min $\theta$ ($'$) for {\sf source\_mode\sc=uniform}\\
& {\sf source\_rmax} &	max $\theta$ ($'$) for {\sf source\_mode\sc=uniform}\\
& {\sf region\_mode} &	{\sc skyfits,detfits,skyreg,detreg}\\
& {\sf num\_region} &	number of accumulation regions \\%(1--200)\\
& {\sf regfile$N$} &	region file {\#}$N$, $N$=1$\;\sim\;${\sf num\_region}\\
& {\sf arffile$N$} &	output ARF {\#}$N$, $N$=1$\;\sim\;${\sf num\_region}\\
& {\sf detmask} &	mask image in DET coord.\ or {\sc none} \\
& {\sf limit\_mode} &	{\sc mixed,num\_photon,accuracy} \\
& {\sf num\_photon} &	number of photons for each energy \\
& {\sf accuracy} &	calculation accuracy for each energy \\
& {\sf gtifile} &	name of the GTI file or {\sc none} \\
& {\sf date\_obs} &	date of observation for {\sf gtifile\sc=none}\\
& {\sf attitude} &	name of the attitude file or {\sc none} \\
& {\sf ea1}, {\sf ea2}, {\sf ea3} &
			Euler angles for {\sf attitude\sc=none} \\
& {\sf rmffile} &	RMF to retrieve energy bin \\
& {\sf estepfile} &	$E$ step file or {\sc dense,medium,sparse}\\
& ({\sf contamifile}) &	XIS contamination file or {\sc none}\\
& ({\sf aberration}) &	enable the aberration correction \\
& ({\sf aperture\_cosine})\hspace*{-1em}&
			consider aperture decrease by $\cos\theta$ \\
\makebox[0in][l]{SimASTE\_XRTsim $^\dagger$} \\
\makebox[0in][l]{SimASTE\_XRTOUTtoDET} \\
\makebox[0in][l]{SimASTE\_XISarfBuild} \\
\hline\\[-1ex]
\multicolumn{3}{l}{\parbox{0.47\textwidth}{\footnotesize
\footnotemark[$\dagger$]
See table~\ref{tab:xissim} for other parameters.}}\\
\end{tabular}
}
\end{table}

\subsection{xissimarfgen}
\label{subsec:xissimarfgen-parameter}

Table~\ref{tab:xissimarfgen} summarizes the structure and
parameters of {\sl xissimarfgen}.
It consists of five ANL modules, and three out of which are common to
{\sl xissim}. The two dedicated modules for {\sl xissimarfgen} are
XISarfPhotonGen and XISarfBuild, and they closely
cooperate to calculate and generate the resultant ARF(s) by driving
the XRT part of the simulator, XRTsim and XRTOUTtoDET\@.

In table~\ref{tab:xissimarfgen},
parameters of common modules to {\sl xissim} are omitted,
although the {\sf simulation\_mode} parameter is set to {\sc weight}
as mentioned in A.\ref{subsec:xissim-structure}.
We can categorize them as follows:
(a) to specify the spatial distribution of the celestial
target on the sky, {\sf source\_mode}, {\sf source\_image}, etc;
(b) to specify the accumulation region of the detected events
and corresponding output ARF names,
{\sf region\_mode}, {\sf num\_region}, {\sf regfile$N$},
{\sf detmask}, and {\sf arffile$N$};
(c) to specify the photon statistics at each energy,
{\sf limit\_mode}, {\sf num\_photon}, and {\sf accuracy};
(d) to specify the energy step to calculate the detection efficiency,
{\sf rmffile} and {\sf estepfile};
(e) to specify the observation date and the satellite Euler angles,
{\sf gtifile}, {\sf date\_obs}, {\sf attitude},
{\sf ea1}, {\sf ea2}, and {\sf ea3};
(f) to specify other calibration information or simulation modes
or reference of the SKY coordinate,
{\sf contamifile}, {\sf aberration}, {\sf aperture\_cosine},
{\sf pointing}, {\sf ref\_alpha}, {\sf ref\_delta}, and {\sf ref\_roll}.
Groups (e) and (f) parameters are similar to {\sl xissim}.

Group (a) parameters determine the spatial distribution of the target
on the sky, and one can specify an arbitrary FITS image in the SKY
or DET coordinate ({\sf source\_mode\sc=skyfits/detfits}).
Pixels with negative values are treated as zero in the image.
Otherwise, a location of a point source can
be set in the equatorial coordinate in J2000, or SKY- or DET-coordinate
({\sf source\_mode\sc=j2000/skyxy/detxy}).
In addition, a uniform-sky emission with respect to the XRT coordinate
can be selected ({\sf source\_mode\sc=uniform}).
When the FITS image or the location of the point source is supplied
in the DET coordinate, its position on the sky will be affected by
the wobbling of the spacecraft, hence it is not recommended
to use with the attitude file.

Note that one must specify {\it skyref}\/ to use SKY coordinates.
When {\sf source\_mode\sc=skyfits}, {\it skyref}\/ is automatically
read from the FITS header keywords. As long as the WCS
(world coordinate system; \cite{Greisen2002,Calabretta2002})
keywords are correctly assigned, one may use an image for {\sf
  source\_image}. 
When {\sf source\_mode\sc=skyxy}, things are a little complicated.
If {\sf pointing\sc=user}, the {\sf ref\_alpha}, {\sf ref\_delta},
and {\sf ref\_roll} parameters are utilized for {\it skyref}.
If {\sf pointing\sc=auto}, which is the default,
{\it skyref}\/ is read from the header keywords of the attitude file,
{\sc ra\_nom} and {\sc dec\_nom}, 
and {\sc roll} of {\it skyref}\/ is always set to 0$^\circ$,
unless {\sf attitude\sc=none}.
If {\sf pointing\sc=auto} and {\sf attitude\sc=none},
{\it skyref}\/ is calculated from the specified Euler angles,
{\sf ea1}, {\sf ea2}, and {\sf ea3},
as $\makebox{\sc ra}=\makebox{\sf ea1}$,
$\makebox{\sc dec}=90^\circ - \makebox{\sf ea2}$, and
$\makebox{\sc roll}=0^\circ$.

Group (b) parameters decide the accumulation region(s) of
the detected events. The {\sf num\_region} parameter specify
the number of regions to be considered in the ARF calculation.
Up to 200 regions may be specified in a single batch of simulations.
One may specify FITS image(s) or DS9-style region file(s)
in SKY- or DET-coordinate
({\sf region\_mode\sc=sky\-fits/det\-fits/sky\-reg/det\-reg})\@.
In supplying a FITS image, an unbinned image ($1536\times 1536$ for SKY,
$1024\times 1024$ for DET) is needed to avoid ambiguity.
The {\it skyref}\/ is adopted in the same way
when {\sf source\_mode\sc=skyxy},
and the header keywords in the FITS image(s) are always ignored.
One may optionally set the {\sf detmask} parameter
to specify a mask image in DET or ACT coordinates,
which is automatically judged by the {\sc ctype1} and {\sc ctype2}
header keywords. The {\sf detmask} image is commonly applied to
the all of the specified regions, so that this feature is useful in
excluding the calibration source regions, bad CCD columns,
and hot/flickering pixels from the accumulation regions.

In supplying a FITS image with the {\sf regfile$N$} parameter,
the pixel values are interpreted as follows.
After the simulation of each photon, the pixel location
on the image is determined.
If the pixel value is zero or negative, the photon is discarded as
a non-detection. If it is positive, the {\sc weight} of the photon is
multiplied by the pixel value. Therefore, one should normally supply a binary
(0/1) mask image, while a gray-scale image can represent non-uniform
exposure areas.

Group (c) parameters specify conditions for the photon statistics
at each simulation energy.
When {\sf limit\_mode\sc=num\_photon}, {\sf num\_photon} count of
photons are generated at each simulation energy,
regardless of the number of detected photons.
When {\sf limit\_mode\sc=accuracy}, photons are generated until
the relative error of the detection efficiency becomes less than
the {\sf accuracy} parameter for all the accumulation regions.
The relative error is calculated by eq.~(\ref{eq:relerr}).
It may take so many photons when the detection efficiency is quite low
in this mode that it is recommended to specify {\sf limit\_mode\sc=mixed},
in which it generates photons until either of the two conditions is fulfilled.
If {\sf num\_photon}=0 or {\sf accuracy}=0 in {\sf limit\_mode\sc=mixed},
the former or the latter is ignored, respectively.

Group (e) parameters are utilized to specify the simulation energy step.
The XIS RMF specified by the {\sf rmffile} parameter is only used
to retrieve the energy bin, $E_i$, for the output ARF(s),
so that an out-of-date RMF may do so far as the energy bin is the same.
The {\sf estepfile} parameter should be one of the preset keywords
{\sc full/dense/medium/sparse}, or point to a file that contains
three decimal numbers, $E_{\rm min}$, $E_{\rm max}$, $E_{\rm bin}$,
on each line. When {\sf estepfile\sc=full}, the simulation is conducted
at every RMF energy bin ($m=7900$ energy steps), randomizing each photon
energy within a $\pm 1$~eV range, which can result in a very long
computation time.
When {\sf estepfile\sc=dense/medium/sparse}, the simulation is conducted
at built-in fixed energies of 2303/157/55 steps, respectively.
These energies are optimized considering the edge energies,
so that even {\sf estepfile\sc=sparse} can produce an acceptable
quality ARF for the scientific spectral fitting analysis with
a featureless and moderate-statistics spectrum.
%One can arrange his/her own energy step by supplying a list of
%$E_{\rm min}$, $E_{\rm max}$, $E_{\rm bin}$ in a text file.

\section{Output File Formats}\label{app:file-formats}

\begin{table}
\caption{
List of columns in the photon file.
}\label{tab:photon-file}
\centerline{\small
\begin{tabular}{llll}\hline\hline
Column Name & \makebox[0in][c]{Format$^*$} & Unit & Description \hspace*{6em}\\
\hline
{\sc photon\_time}	&{\sc 1d}& s	& arrival time \\
{\sc photon\_energy}	&{\sc 1e}& keV	& X-ray energy \\
{\sc ra}		&{\sc 1e}& deg	& right ascension of incidence \\
{\sc dec}		&{\sc 1e}& deg	& declination of incidence \\
\hline\\[-1ex]
\multicolumn{4}{l}{\parbox{0.47\textwidth}{\footnotesize
\footnotemark[$*$] The column format in FITS convention.
`{\sc 1d}' is a double precision floating point,
and `{\sc 1e}' is a single precision floating point.}}
\end{tabular}
}
\end{table}

\begin{table}
\caption{
List of columns in the simulated event file.
}\label{tab:event-file}
\centerline{\small
\begin{tabular}{llll}\hline\hline
Column Name & \makebox[0in][c]{Format$^*$} & Unit & Description \hspace*{6em}\\
\hline
\multicolumn{4}{l}{\sc 1st extension {\sc 'events'}}\\
{\sc time}		&{\sc 1d}& s		& detected time \\
{\sc pha}		&{\sc 1i}& chan		& pulse height (= {\sc pi}) \\
{\sc pi}		&{\sc 1i}& chan		& pulse invariant \\
{\sc status}		&{\sc 1i}&		& status flags (= 0) \\
{\sc grade}		&{\sc 1i}& 		& event grade (= 0) \\
{\sc segment}		&{\sc 1i}&		& CCD segment id \\
{\sc rawx}		&{\sc 1i}& pixel	& RAW coordinate x value \\
{\sc rawy}		&{\sc 1i}& pixel	& RAW coordinate y value \\
{\sc actx}		&{\sc 1i}& pixel	& ACT coordinate x value \\
{\sc acty}		&{\sc 1i}& pixel	& ACT coordinate y value \\
{\sc detx}		&{\sc 1i}& pixel	& DET coordinate x value \\
{\sc dety}		&{\sc 1i}& pixel	& DET coordinate y value \\
{\sc focx}		&{\sc 1i}& pixel	& FOC coordinate x value \\
{\sc focy}		&{\sc 1i}& pixel	& FOC coordinate y value \\
{\sc x}			&{\sc 1i}& pixel	& SKY coordinate x value \\
{\sc y}			&{\sc 1i}& pixel	& SKY coordinate y value \\
{\sc xisx}		&{\sc 1e}& pixel
	& floating point value of {\sc detx}\hspace*{-2em}\\
{\sc xisy}		&{\sc 1e}& pixel
	& floating point value of {\sc dety}\hspace*{-2em}\\
{\sc weight}		&{\sc 1e}&		& weight of the event \\
{\sc photon\_time}	&{\sc 1d}& s		& copy of input photon file \\
{\sc photon\_energy}	&{\sc 1d}& keV		& copy of input photon file \\
{\sc ra	}		&{\sc 1d}& deg		& copy of input photon file \\
{\sc dec}		&{\sc 1d}& deg		& copy of input photon file \\
\hline
\multicolumn{4}{l}{\sc 2nd extension {\sc 'gti'}}\\
{\sc start}		&{\sc 1d}& s		& start of valid time \\
{\sc stop}		&{\sc 1d}& s		& stop of valid time \\
\hline\\[-1ex]
\multicolumn{4}{l}{\parbox{0.47\textwidth}{\footnotesize
\footnotemark[$*$] The column format in FITS convention.
`{\sc 1d}' is a double precision floating point,
`{\sc 1e}' is a single precision floating point,
`{\sc 1i}' is a 16-bit integer.
}}
\end{tabular}
}
\end{table}

\begin{table}
\caption{
List of columns in the generated ARF\@.
}\label{tab:arf-columns}
\centerline{\small
\begin{tabular}{llll}\hline\hline
Column Name & \makebox[0in][c]{Format$^*$} & \makebox[0in][l]{Unit} & Description \hspace*{6em}\\
\hline
\multicolumn{4}{l}{\sc Primary extension {\sc 'wmap'}}\\
$[$image$]$		&\makebox[0.8em][l]{\sc 1d}& \makebox[0in][l]{count}
	& image of detected events \\
\hline
\multicolumn{4}{l}{\sc 1st extension {\sc 'specresp'}}\\
{\sc energ\_lo}		&\makebox[0in][l]{\sc 1e}& keV	& lower energy bin \\
{\sc energ\_hi}		&\makebox[0in][l]{\sc 1e}& keV	& higher energy bin \\
{\sc specresp}		&\makebox[0in][l]{\sc 1e}&cm$^2$& computed effective area \\
{\sc resperr}		&\makebox[0in][l]{\sc 1e}&cm$^2$& error of {\sc specresp} \\
{\sc resprerr}		&\makebox[0in][l]{\sc 1e}& 	& relative error of {\sc specresp} \\
{\sc xrt\_effarea}	&\makebox[0in][l]{\sc 1e}&cm$^2$& XRT only effective area \\
{\sc shield\_transmis}	&\makebox[0in][l]{\sc 1e}&	& \makebox[0in][l]{thermal shield transmission} \\
{\sc contami\_transmis}\hspace*{-1em}
			&\makebox[0in][l]{\sc 1e}&	& \makebox[0in][l]{contamination transmission} \\
{\sc index}		&\makebox[0in][l]{\sc 1j}&
	& index of simulated energy\hspace*{-2em}\\
{\sc s}			&\makebox[0in][l]{\sc 1e}&	& interpolation coefficient \\
{\sc t}			&\makebox[0in][l]{\sc 1e}&	& interpolation coefficient \\
{\sc input}		&\makebox[0in][l]{\sc 1e}& \makebox[0in][l]{count}
	& number of input photons \\
{\sc detect}		&\makebox[0in][l]{\sc 1e}& \makebox[0in][l]{count}
	& sum of detected events \\
{\sc weisum}		&\makebox[0in][l]{\sc 1e}& \makebox[0in][l]{count}
	& weighted sum of events\hspace*{-2em}\\
{\sc relerr}		&\makebox[0in][l]{\sc 1e}& 	& relative error of {\sc detect} \\
\hline\\[-1ex]
\multicolumn{4}{l}{\parbox{0.47\textwidth}{\footnotesize
\footnotemark[$*$] The column format in FITS convention.
`{\sc 1d}' is a double precision floating point,
`{\sc 1e}' is a single precision floating point,
`{\sc 1j}' is a 32-bit integer.
}}
\end{tabular}
}
\end{table}

\begin{table}
\caption{
\makebox{List of special header keywords in the output ARF.}
}\label{tab:arf-header-keys}
\centerline{\small
\begin{tabular}{ll}\hline\hline
Keyword Name & Description \hspace*{6em}\\
\hline
{\sc geomarea} 		& geometrical area of XRT (cm$^2$) \\
{\sc teldef}		& {\it teldef}\/ file name \\
{\sc leapfile}		& leap second file name \\
{\sc creator}		& {\sl xissimarfgen} credit and version \\
{\sc source\_ratio\_reg}\hspace*{-0.8em}
	& {\sf source\_image} ratio inside selected region\hspace*{-0.5em} \\
{\sc mask\_ratio\_ccd}	& {\sf detmask} ratio in the whole CCD area \\
{\sc mask\_ratio\_reg}	& {\sf detmask} ratio inside selected region \\
{\sc n\_photon}		& number of input photons generated \\
{\sc n\_detect}		& number of events detected $^*$ \\
{\sc n\_weisum}		& weighted sum of events detected $^*$ \\
{\sc randseed}		& random number seed, =\,{\sf rand\_seed} \\
{\sc randskip}		& random number skip count, =\,{\sf rand\_skip}\hspace*{-0.5em} \\
{\sc randngen}		& number of random numbers generated\hspace*{-0.5em} \\
\hline\\[-1ex]
\multicolumn{2}{l}{\parbox{0.47\textwidth}{\footnotesize
\footnotemark[$*$]
Events fallen on a pixel with a positive value for
at least one of the accumulation regions are
treated as ``detection'', here.}}
\end{tabular}
}
\end{table}

\subsection{Output of mkphlist}
\label{subsec:mkphlist-output}

Table~\ref{tab:photon-file} denotes the format of the output photon file
from {\sl mkphlist}, which is also the input to {\sl xissim}.
The time, energy, and direction of the incident photons are
contained in the {\sc photon\_time}, {\sc photon\_energy},
and ({\sc ra}, {\sc dec}) columns, respectivly.
This file is basically mission independent, except for
the {\sc geomarea} keyword in the FITS header,
which contains the geometrical area of XRT (cm$^2$)
specified by the {\sf geometrical\_area} parameter of {\sl mkphlist}.

\subsection{Output of xissim}
\label{subsec:xissim-output}

Table~\ref{tab:event-file} denotes the format of the output event file
from {\sl xissim}.
The output is a standard FITS event file with {\sc events} and {\sc gti}
extensions,
plus information on the faked input photons.
If the attitude file is given, position on the sky ({\sc x} and {\sc y})
is corrected for the wobbling of the spacecraft,
as well as the parallax (aberration).
For the calculation of the {\sc x} and {\sc y} columns,
the sky reference ({\it skyref}\/) of
(${\it RA}_{\rm ref}$, ${\it DEC}_{\rm ref}$, ${\it Roll}_{\rm ref}$)
must be specified by users as the parameters of {\sf ref\_alpha},
{\sf ref\_delta}, and {\sf ref\_roll},
although {\sf ref\_roll} is hidden and usually set to 0.
The ({\sc x}, {\sc y}) positions have a one-to-one correspondence to
sky directions, 
however these direction do not necessarily coincide with the incident
direction of the photon, ({\sc ra}, {\sc dec}), due to blurring
by the PSF of the XRT\@.

\subsection{Output of xissimarfgen}
\label{subsec:xissimarfgen-output}

Table~\ref{tab:arf-columns} summarizes the format of the output ARF
generated by {\sl xissimarfgen}.
In comparison with the minimum set of the ARF
({\sc energ\_lo}, {\sc energ\_hi}, and {\sc specresp} columns),
it has several additional columns and the primary image extension
to record the information of the simulation.
All of the parameters for {\sl xissimarfgen} are written in the
history of the output ARF, and several important simulation parameters
and results are written to the FITS header keywords listed in
table~\ref{tab:arf-header-keys}.

The {\sc energ\_lo} and {\sc energ\_hi} columns,
both in units of keV, are copied from the input RMF,
and {\sl xissimarfgen} assumes
$E_i = (\makebox{\sc energ\_lo} + \makebox{\sc energ\_hi})\; /\; 2$ and 
$\Delta E_i = \makebox{\sc energ\_hi} - \makebox{\sc energ\_lo}$
for the $i$-th row ($i = 1\sim m$, and $m = 7900$ for the nominal RMF)\@.
The {\sc specresp} column holds the final result of
the detection efficiency times the geometrical area,
$S\;A(E_i)$, in unit of cm$^2$. The value of $S$ (cm$^2$) assumed
in the calculation is written to the FITS header keyword of
{\sc geomarea} (table~\ref{tab:arf-header-keys}).
The {\sc resperr} and {\sc resprerr} columns hold
the absolute and relative errors, respectively, estimated for
the {\sc specresp} column,
i.e.\ {\sc resperr} = {\sc resprerr} $\times$ {\sc specresp}.
Details are described in \S\,\ref{subsec:xissimarfgen-implement} for the
{\sc xrt\_effarea}, {\sc shield\_transmis}, 
and {\sc contami\_transmis} columns.

The {\sc input}, {\sc detect}, and {\sc weisum} columns hold interpolated
values of $N_{\rm in}$, $N_{\rm det}$, and $N_{\rm w}$, respectively.
The {\sc weisum} represents the weighted sum of events,
and is usually equal to {\sc detect}
unless one supplies a gray-scale region file.
The {\sc relerr} is an interpolated value of the relative error
calculated by eq.~(\ref{eq:relerr}).
The coefficients of the interpolation, {\sc s}$_i$ and {\sc t}$_i$,
are defined as,
\begin{eqnarray}
V(E_i) &=& \makebox{\sc s}_i\; V(E'_{\makebox{\tiny\sc index}_i})
+ \makebox{\sc t}_i\; V(E'_{\makebox{\tiny\sc index}_i+1}),\\
\makebox{\sc s}_i &=& (E'_{\makebox{\tiny\sc index}_i+1} - E_i)
\label{eq:s}
\;/\; (E'_{\makebox{\tiny\sc index}_i+1} - E'_{\makebox{\tiny\sc index}_i}),\\
\label{eq:t}
\makebox{\sc t}_i &=& (E_i - E'_{\makebox{\tiny\sc index}_i})
\;/\; (E'_{\makebox{\tiny\sc index}_i+1} - E'_{\makebox{\tiny\sc index}_i}),
\end{eqnarray}
where $E'_{\makebox{\tiny\sc index}_i}$ denotes the simulated energy
indexed by an integer value of {\sc index}$_i$,
$V(E_i)$ and $V(E_{\makebox{\tiny\sc index}_i})$ denotes
a value at the $i$-th row and the simulation value at
$E_{\makebox{\tiny\sc index}_i}$~keV,
and {\sc s}$_i$, {\sc t}$_i$, and {\sc index}$_i$ denote
the $i$-th row values of the {\sc s}, {\sc t}, and {\sc index} columns,
respectively. As one can see easily from these formulae,
a simple linear interpolation is adopted in {\sl xissimarfgen}.
Note that these column values in this paragraph do not include
the effect of the XIS contamination, so that {\sc relerr}
is slightly different from {\sc resprerr}.

There is a primary extension image written in the output ARF, too.
This image holds the collection of all the simulated photons
detected in the specified accumulation region for the ARF calculation.
The coordinates of the image are dependent on the {\sf region\_mode};
SKY coordinate ($1536\times 1536$) for {\sf region\_mode\sc=sky*}, and
DET coordinate ($1024\times 1024$) for {\sf region\_mode\sc=det*}.
The {\sc weight} value of each photon without contamination
is filled in the image, so that the image usually contains
integer pixel values, unless one supplies an gray-scale region file.
Pixels out of the accumulation region are filled with a value of $-1$,
which is useful in checking that the accumulation region is correctly
assigned. One can also examine whether the celestial target is
correctly placed in the FOV\@.

\end{document}